\documentclass[a4paper,11pt]{article}
\pdfoutput=1 

\usepackage{jheppub} 

\usepackage{amsmath, amsthm, amssymb}
\usepackage{dsfont}
\usepackage{wasysym, verbatim} 
\usepackage{color}
\usepackage{tensor}
\usepackage{mathtools}
\usepackage{enumitem}
\usepackage{physics}
\usepackage{subcaption}
\usepackage{graphicx}
\usepackage[utf8]{inputenc}
\usepackage{booktabs,multirow, quotes,tablefootnote}
\usepackage{nicematrix}
\usepackage{tikz}
\NiceMatrixOptions
{
	custom-line = 
	{
		letter = : ,
		command = dashedline , 
		ccommand = cdashedline ,
		tikz = dashed
	},
	custom-line = 
	{
		letter = . ,
		command = dotrule , 
		ccommand = cdotrule ,
		tikz = dotted
	}
}

\usepackage{cancel}
\usepackage[normalem]{ulem}

\usepackage{braket}
\usepackage{tcolorbox}
\usepackage{float}

\usepackage{notoccite} 

\usepackage{cleveref}
\creflabelformat{equation}{(#2\textup{#1}#3)}
\crefname{equation}{Eq.}{Eqns.}
\crefname{figure}{Fig.}{Figs.}
\crefname{section}{Section}{Sections}
\crefname{table}{Table}{Tables}
\crefname{chapter}{Chapter}{Chapters}
\crefname{appendix}{Appendix}{Appendices}
\crefname{subsection}{Section}{Sections}
\crefname{remark}{Remark}{Remarks}
\crefname{footnote}{footnote}{footnotes}

\usepackage{cases}

\usepackage{xfrac}
\newcommand{\oneeighth}{\text{\sfrac 18}}
\usepackage{nicefrac}

\usepackage{xcolor}
\usepackage{hyperref}

\numberwithin{equation}{section}

\theoremstyle{theorem}
\newtheorem*{conjecture*}{Conjecture}

\numberwithin{theorem}{section}
\newtheorem{definition}{Definition}\numberwithin{definition}{section}
\numberwithin{lemma}{section}
\theoremstyle{remark}
\newtheorem{remark}{Remark}\numberwithin{remark}{section}

\usepackage{adjustbox}

\def\eps{\epsilon}
\newcommand{\beq}{\begin{equation}} 
\newcommand{\eeq}{\end{equation}}

\def\calO {{\cal O}}

\def\calD {{\cal D}}

\def\calM {{\cal M}}

\def\calT {{\cal T}}

\def\ge{\geqslant}
\def\le{\leqslant}

\def\<{\langle}
\def\>{\rangle}

\newcommand{\myinclude}[2][]{\raisebox{0.6ex}{\raisebox{-0.5\height}{\includegraphics[#1]{#2}}}}

\usepackage{enumitem}

\renewcommand{\ge}{\geqslant}
\renewcommand{\le}{\leqslant}

\makeatletter
\def\@fpheader{\ }
\makeatother

\title{Transfer Matrix and Lattice Dilatation Operator\\ for High-Quality Fixed Points\\ in Tensor Network Renormalization Group}

\author[a]{Nikolay Ebel,}
\author[b]{Tom Kennedy}
\author[a]{and Slava Rychkov}

\affiliation[a]{Institut des Hautes \'Etudes Scientifiques, 91440 Bures-sur-Yvette, France}
\affiliation[b]{Department of Mathematics, University of Arizona,
	Tucson, AZ 85721, USA}

\emailAdd{ebelnikola@gmail.com}
\emailAdd{slava@ihes.fr}
\emailAdd{tgk@arizona.edu}

\abstract{
  Tensor network renormalization group maps study critical points of 2d lattice models like the Ising model by finding the fixed point of the RG map. In a prior work \cite{Ebel:2024nof} we showed that by adding a rotation to the RG map, the Newton method could be implemented to find an extremely accurate fixed point. For a particular RG map (Gilt-TNR) we studied the spectrum of the Jacobian of the RG map at the fixed point and found good agreement between the eigenvalues corresponding to relevant and marginal operators and their known exact values. In this companion work we use two further methods to extract many more scaling dimensions from this Newton method fixed point, and compare the numerical results with the predictions of conformal field theory (CFT). The first method is the well-known transfer matrix (TM), while the second method we refer to as the lattice dilatation operator (LDO). We introduce some extensions of these method that provide also spins of the CFT operators, modulo an integer. With comparable computing resources, the TM and LDO methods perform equally well. The agreement for the scaling dimensions and spins is excellent up to $\Delta=4\oneeighth$, and reasonably good up to 2 units higher. Some of the eigenvalues of the Jacobian of the RG map can come from perturbations associated with total derivative interactions and so are not universal. In some past studies \cite{Lyu:2021qlw,PhysRevE.109.034111} such non-universal eigenvalues did not appear in the Jacobian. We explain this surprising result by showing that their RG map has the unusual property that the Jacobian is equivalent to the LDO operator. 
}

\begin{document}

\maketitle

\flushbottom

\section{Introduction}

Renormalization group (RG) algorithms using tensor networks ("tensor RG") \cite{Levin:2006jai,SRG,TEFR,SRG1,HOTRG,Evenbly-Vidal,LoopTNR,Bal:2017mht,Evenbly:2017dyd,GILT,Ebel:2025ezv} are a promising alternative to the more traditional Wilson-Kadanoff block-spin techniques \cite{Wilson:1974mb,DombGreenVol6Niemeijer} for performing renormalization of two-dimensional (2D) lattice models \cite{Kadanoff}.\footnote{Both Wilson-Kadanoff RG and tensor RG compute a map in a space of couplings, and look for a fixed point. In the former case the couplings are coefficients of interaction terms in a Hamiltonian, while in the latter case the couplings are elements of a tensor. We refer to \cite{paper1} for a detailed description of similarities and differences. Another technique which is further away is Monte Carlo RG \cite{PhysRevLett.42.859}, which uses Monte Carlo simulations to evaluate the Jacobian of a Wilson-Kadanoff block spin transformation at a fixed point, without computing the map itself nor the fixed point. }
	
The key object of study of tensor RG is the fixed point tensor associated with the critical point of the lattice model of interest. Traditionally, the fixed point tensor is found by the "shooting" method, which iterates the RG map from the initial tensor on the critical manifold. In a recent work \cite{Ebel:2024nof} we proposed a new way to look for the fixed point tensor, which solves the fixed point equation via the Newton method. In this way, we were able to find the fixed point tensor with unprecedented accuracy of $10^{-9}$ in the Hilbert-Schmidt norm.
  
Physical information about the critical point, such as the critical exponents, can be extracted if the fixed point tensor is known. In \cite{Ebel:2024nof}, we used the eigenvalues of the Jacobian of the RG map at the fixed point to study the critical exponents. This is a time-honored general method going back to the early days of RG \cite{Wilson:1973jj,DombGreenVol6-Wegner}. But before our work \cite{Ebel:2024nof} it was used only a couple of times in the context of tensor RG, in Ref.~\cite{Lyu:2021qlw,PhysRevE.109.034111}. This method was particularly natural in \cite{Ebel:2024nof} since the Jacobian is also the main ingredient of the Newton method. One limitation of the Jacobian is that the eigenvalues of "redundant" perturbations associated with total derivative interactions are not universal \cite{DombGreenVol6-Wegner} and cannot be meaningfully compared with the scaling dimensions of the corresponding operators in conformal field theory (CFT).

Instead of the Jacobian, most prior tensor RG works extracted the critical exponents using two other methods: the transfer matrix (TM) \cite{TEFR} and the lattice dilatation operator (LDO) \cite{TNRScaleTrans}.\footnote{Unfortunately a universally accepted term is still lacking. Previous works used "scaling map," "scale transfer matrix", "transfer matrix on a cylinder" \cite{TNRScaleTrans}, "radial transfer matrix MPO" \cite{Bal:2017mht}, "ascending superoperator" \cite{Evenbly-review}, "super operator" \cite{IinoMoritoKawashima}, and "scaling operator from logarithmic transformation" \cite{Evenbly-TNR-website}. We propose "lattice dilatation operator", as we believe this term best captures its essence, better distinguishes it from the transfer matrix, and avoids confusion with the standard meaning that ``scaling operator'' has in RG theory.} The TM and LDO are methods specific to tensor RG, taking advantage of the locality properties of tensor networks. Indeed, it may be impossible to set up something like TM or LDO for other kinds of RG for which the fixed-point Hamiltonian is expected to be non-local, such as e.g.~the Wilson-Kadanoff block-spin transformation RG \cite{Wilson:1974mb,Tom}.

It is interesting to investigate how the TM and LDO methods perform when applied to the high-quality fixed point tensors obtained in \cite{Ebel:2024nof} using the Newton method. This is the main purpose of this paper.

This paper is structured as follows. In \cite{Ebel:2024nof} we showed how the Newton method could be implemented to find an extremely accurate fixed point of a tensor RG map. This required adding a rotation to the RG map. We review those results in \cref{sec:brief}. In \cref{sec:transfer} we consider the TM method to extract scaling dimensions from the RG fixed point, and we introduce some extensions of this method that provide spins of the CFT operators modulo an integer. We give numerical results for the Gilt-TNR RG map. \cref{sec:DSO} explains the theory of the LDO method for extracting scaling dimensions as well as spins modulo 4, constructs the LDO operator for Gilt-TNR, and gives numerical results for this RG map. Conclusions and questions for further study are in \cref{sec:conclusions}. 

The appendices provide more details on the TM and LDO methods. In particular, in \cref{app:frozenJacs} we discuss the connection between the LDO and the Jacobians in Refs.~\cite{Lyu:2021qlw,PhysRevE.109.034111}. There, we explain the surprising agreement between Jacobian eigenvalues of the RG maps from \cite{Lyu:2021qlw,PhysRevE.109.034111} with the CFT scaling dimensions of derivative operators. This agreement is surprising as the eigenvalues corresponding to total derivative operators are not universal \cite{Ebel:2024nof}. We show that it can be explained by the coincidence of the Jacobians in \cite{Lyu:2021qlw,PhysRevE.109.034111} with the LDO operators for the corresponding RG steps. We note that such a coincidence is not a common feature of tensor RG maps.

\section{Recap of tensor RG and of \cite{Ebel:2024nof}}
\label{sec:brief}

This paper is a companion work to \cite{Ebel:2024nof}, but in an effort to make it self-contained we will review here the main aspects of tensor RG and of the results of \cite{Ebel:2024nof}.

\subsection{Tensor networks}
\label{sec:TN}
\begin{figure}
	\centering
	\includegraphics[scale=0.6]{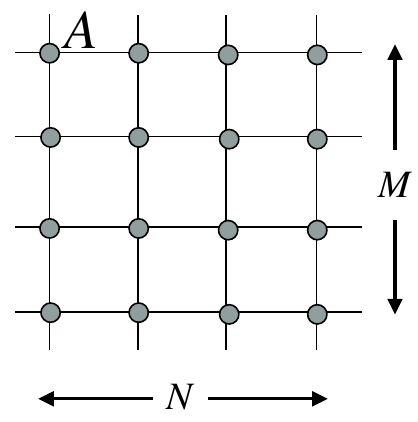}
	\caption{$N\times M$ tensor network built out of a four-legged tensor $A$. We are assuming periodic boundary conditions, so that the outgoing bonds on network sides are contracted pairwise.
		\label{fig:TN} 
	}
\end{figure}
 
In the tensor network approach to RG one represents the partition function of a 2D lattice model as a contraction of a tensor network made out of a four-legged tensor $A$, \cref{fig:TN}. The partition function of such a $N\times M$ tensor network is denoted by $Z(A,N\times M)$. One important parameter of the tensor network is the bond dimension $\chi$, which is the number of values that every tensor index can take. Numerical algorithms necessarily work with a finite $\chi$. Sometimes one needs to discuss tensors of infinite $\chi$. For example, the exact critical fixed point tensor (see \cref{sec:shooting}) is expected to have $\chi=\infty$. Infinite-bond-dimension tensors are OK as long as they have finite Hilbert-Schmidt norm, defined as
\beq
\|A\|^2 = \sum_{ijkl}|A_{ijkl}|^2\,.
\eeq
For example, the partition function $Z(A,N\times M)$ is then finite \cite[Prop.~2.3]{paper2}.

\subsection{Tensor RG}
\label{sec:TRG}

An RG map $\mathcal{R}:A\to A'$ preserves the partition function while simultaneously reducing the dimensions of the network by a constant integer factor $b>1$:\footnote{We note that a single step of the TRG map on square lattice, \cite[Fig.~3]{Levin:2006jai}, would not fit this definition as it rotates the lattice by $\pi/4$ and reduces the dimensions by $\sqrt{2}$. Two steps of TRG would be OK, with $b=2$.} 
\beq
Z(A,N\times M)=Z(\mathcal{R}(A),N/b\times M/b)\,.
\label{eq:nonrotZ}
\eeq
It is important that this equation holds for any $N,M$ divisible by $b$, and that $\mathcal{R}(A)$ does not depend on $N,M$. Below we will work with the map $\mathcal{R}(A)$, while the arbitrary lengths $N,M$ will not play any role.\footnote{In particular $N,M$ will not influence the fixed point equations or their approximations.}

The map $\mathcal{R}$ is called non-normalized in the sense that $\|\mathcal{R}(A)\|\ne1$ in general. The normalized RG map $R$ is instead defined by rescaling $\mathcal{R}(A)$ by a constant to get a tensor of unit Hilbert-Schmidt norm:
\beq
R(A)=\|\mathcal{R}(A)\|^{-1}\mathcal{R}(A)\,.
\eeq
The normalized RG map preserves the partition function up to rescaling by a constant factor $\|\mathcal{R}(A)\|^{NM/b^2}$. 

\begin{figure}
	\centering
	\includegraphics[scale=0.6]{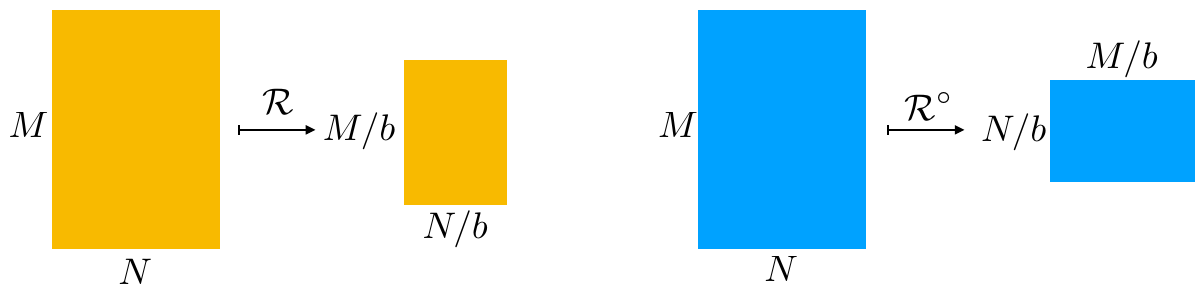}
	\caption{A non-rotating RG map (left) vs a rotating RG map (right)
		\label{fig:nonrot-vs-rot}
	}
\end{figure}

The RG maps $\mathcal{R}$ and $R$ above are called non-rotating because they preserve the orientation of the lattice. The associated rotating RG map $\mathcal{R}^\circ$ is defined \cite{Ebel:2024nof} by composing $\mathcal{R}$ with a $\pi/2$ rotation $\Gamma_{\pi/2}$:
\beq
\mathcal{R}^\circ = \Gamma_{\pi/2} \mathcal{R}\,,
\label{eq:Rwithrot}
\eeq
or graphically:
\beq
\myinclude[scale=0.7]{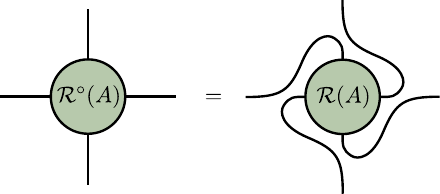}\,
\label{eq:RGwithrot}
\eeq
(and analogously $R^\circ$).
So, the rotating map $\mathcal{R}^\circ$ maps a tensor network of size $N\times M$ to one which has size $M/b\times N/b$, \cref{fig:nonrot-vs-rot}, and Eq.~\eqref{eq:nonrotZ} is modified accordingly:
\beq
Z(A,N\times M)=Z(\mathcal{R}^\circ(A),M/b\times N/b)\,.
\label{eq:rotZ}
\eeq

The above discussion was for an exact tensor RG map. In practical calculations, tensors $A$ and $A'$ related by an RG map are represented as numerical tensors of the same finite bond dimension $\chi$, depending on the available computer resources. It is then usually impossible to satisfy Eq.~\eqref{eq:nonrotZ} exactly, but only approximately. Different tensor RG algorithms \cite{Levin:2006jai,SRG,TEFR,SRG1,HOTRG,Evenbly-Vidal,LoopTNR,Bal:2017mht,Evenbly:2017dyd,GILT,Ebel:2025ezv} have been devised to minimize the truncation error in satisfying this equation. Typically, successful tensor RG algorithms involve a combination of steps referred to as coarse-graining, gauge-fixing, and disentangling (or entanglement filtering) \cite{TEFR,Evenbly-Vidal}. Examples of such algorithms include TNR \cite{Evenbly-Vidal}, Loop-TNR \cite{LoopTNR}, TNR${}_{+}$ \cite{Bal:2017mht}, Gilt-TNR \cite{GILT}, 2x1 \cite{Ebel:2025ezv}. Inclusion of the disentangling step makes such algorithms significantly more complicated than the original TRG algorithm \cite{Levin:2006jai}, but it is necessary to solve the so called corner-double-line (CDL) tensor problem and reach a fixed point \cite{Levin-talk}. In this paper as in \cite{Ebel:2024nof} we will use the Gilt-TNR algorithm \cite{GILT}, reviewed in \cite[App.~C]{Ebel:2024nof}. This algorithm is differentiable in the neighborhood of the fixed point, as we verified in \cite{Ebel:2024nof}. This will be important for us to implement the Newton method. Any other algorithm with disentangling which is differentiable would also be suitable.

\subsection{Critical fixed point via ``shooting'' method}
\label{sec:shooting}

The partition function of any statistical physics lattice model with a finite-range interaction Hamiltonian can be transformed into tensor network form \cite[Prop.~3.2]{paper2}. Consider the nearest-neighbor 2D Ising model. We focus on the isotropic case of equal horizontal and vertical couplings (in \cite{Ebel:2024nof} the anisotropic case was also considered). The resulting tensor $A=A_{\rm NN}(t)$ has bond dimension 2 and is parametrized by the (reduced) temperature $t$, where $t=t_c=1$ corresponds to the critical point.\footnote{Transformation of the partition function to the tensor network form can be done in several inequivalent ways. Here as in \cite{Ebel:2024nof} we use a popular method \cite{TEFR} which involves rotating the lattice of Ising spins by $\pi/4$ \cite[App.~B]{Ebel:2024nof}.\label{ftnt:Ainit}} 

For a suitable RG map $R$, the sequence of normalized RG iterations 
\beq
A^{(n+1)}=R(A^{(n)})\quad(n=0,1,2,\ldots),
\label{eq:RGiters}
\eeq
with
\beq
A^{(0)}=A_{\rm NN}(t),
\eeq
is expected to converge to the high-temperature fixed point for $t>1$, to the low-temperature fixed point for $t<1$ and to the critical fixed point $A_*$ for $t=1$. "Suitable" means that care should be taken to filter out redundant network perturbations known as corner-double-line (CDL) tensors \cite{Levin-talk}. Available numerical tensor RG algorithms working at finite $\chi$, including Gilt-TNR, accomplish this by a disentangling step \cite{Evenbly-Vidal}, whose precise implementation depends on the algorithm.
 
The high- and low-temperature fixed points are simple finite bond dimension tensors. Recently, tensor RG near them was understood in an exact setting not involving any truncation \cite{paper1,paper2}. Here we focus on the critical fixed point tensor $A_*$. In an exact setting, $A_*$ is expected to be a Hilbert-Schmidt tensor with an infinite bond dimension. The rigorous theory of $A_*$ is not yet available. In particular, a tensor RG map capable of solving the CDL problem in the infinite bond dimension setting has not been defined. 

The best we can currently do is to apply a numerical tensor RG algorithm working at a finite bond dimension $\chi$ and find its fixed point, which we will denote $A^{[\chi]}_*$. This fixed point will depend not only on $\chi$ but also on the precise tensor RG algorithm used, and on whatever other parameters the algorithm may have, such as $\epsilon_{\rm gilt}$ in case of Gilt-TNR; we leave these other dependencies implicit. We hope that the fixed point tensor $A^{[\chi]}_*$ provides a good approximation to the as-yet-unknown exact critical fixed point $A_*$ of an as-yet-unknown exact RG map, so we would like to find it. 

One way to find $A^{[\chi]}_*$ is by iterations \eqref{eq:RGiters} starting from the critical nearest-neighbor tensor $A_{\rm NN}(t=1)$. We refer to this as the "shooting" method. It was used in all tensor RG studies before our work \cite{Ebel:2024nof}. For an RG map differentiable at the fixed point, we expect the error to decrease exponentially
\beq
\| A^{(n)}-A^{[\chi]}_*\|\sim (\lambda_{\rm irr})^n\,,
\label{eq:distdicr1}
\eeq
where $\lambda_{\rm irr}<1$ is the leading irrelevant $\mathbb{Z}_2$-even eigenvalue of the Jacobian $\nabla R$ at the fixed point.\footnote{\label{note:Z2even}Only $\mathbb{Z}_2$-even eigenvalues matter because we are assuming that the RG map preserves the $\mathbb{Z}_2$ spin flip symmetry of the Ising model, so that the whole RG evolution happens within the space of $\mathbb{Z}_2$-even tensors.} After $n_*$ iterations, we achieve $(\lambda_{\rm irr})^{n_*}$ accuracy. We also expect the distance between consecutive iterates to decrease exponentially: 
\beq
\| A^{(n+1)}-A^{(n)}\|\sim (\lambda_{\rm irr})^n\,.
\label{eq:distdicr}
\eeq

There are a couple of complications that have to be taken care of when implementing the "shooting" method. The first complication has to do with gauge freedom. The tensor network partition function is invariant if the horizontal and vertical legs of tensor $A$ are multiplied by orthogonal\footnote{In this paper we only work with real tensors.} tensors $G_h$ and $G_v$ and their transposes:
\begin{equation}\label{eq:GaugeFixing1}
	\myinclude[scale=1]{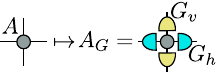}.
\end{equation}
If this gauge freedom is not fixed, the RG map iterates may not approach a single fixed point but a manifold of gauge-equivalent tensors. Thus any algorithm wishing to exhibit \eqref{eq:distdicr} needs to implement a gauge fixing procedure. This was implemented and discussed at length in \cite[App.~D]{Ebel:2024nof}.

\begin{figure}
	\centering
	\includegraphics[width=0.6\textwidth]{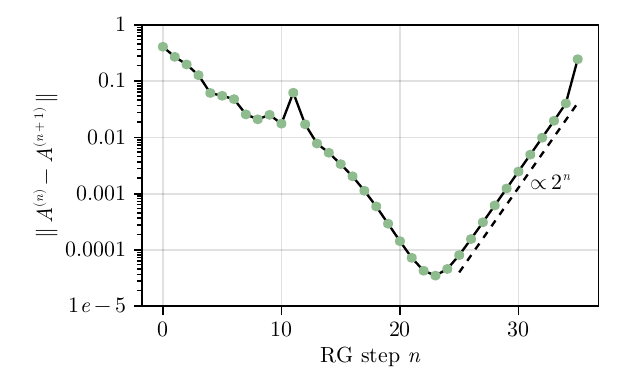}
	\caption{The distance between two consecutive RG iterates as a function of the RG step. The Gilt-TNR parameters are $\chi=30$, $\eps_{\rm gilt}=6\times 10^{-6}$. Figure from \cite{Ebel:2024nof}.
		\label{fig:crit_convergence} 
	}
\end{figure}

The second complication is that the finite-$\chi$ critical temperature $t_c^{\chi}$ is slightly shifted from the exact critical temperature $t_c=1$. We can estimate $t_c^{\chi}$ by the bisection algorithm \cite[Sec.~III~B]{Ebel:2024nof}. Let $\Delta t$ be the width of the final interval in the bisection algorithm. So if we start our iterations at some $t$ in this interval then $|t-t_c^{\chi}| < \Delta t$. Because $t$ is not exactly equal to $t_c^\chi$, the converging behavior \eqref{eq:distdicr} will not continue indefinitely. After a number of steps (depending on $\Delta t$) we will see instead a diverging behavior where the distance grows as $\sim \lambda_{\rm rel}^{n}$, where $\lambda_{\rm rel}=b$ is the leading relevant $\mathbb{Z}_2$-even eigenvalue for the 2D Ising model. The best achievable accuracy $\delta$ of the "shooting" method, i.e.~the minimum over $n$ of the distance $\|A^{(n)}-A_*\|$ from the RG iterates from the fixed point, scales as \cite[Eq.~(3.8)]{Ebel:2024nof}
\beq
\delta\sim \Delta t^\frac{1}{1-1/\log_b \lambda_{\rm irr}}\,.
\label{eq:acc-shooting}
\eeq

An example of this for the Gilt-TNR map with $b=2$, $\chi=30$ is shown in \cref{fig:crit_convergence}. We see a converging behavior with $\lambda_{\rm irr}\sim 0.63$  (the leading irrelevant eigenvalue of $\nabla R$, see \cref{tab:isospec1} below) for $n\lesssim n_*=23$, and a diverging behavior $\sim 2^n$ for larger $n$. The critical temperature is $t_c^\chi= 1.0000110043(1)$. In this figure $A^{(n_*)}$ is the best approximation to the fixed point tensor $A_*^{[30]}$, with accuracy $\delta\sim5\times 10^{-5}$. 

The "shooting" method is general and it would work for the rotating RG map $R^\circ$ just as it did for $R$, up to a small change in the leading irrelevant eigenvalue which controls the rate of approach to the fixed point ($\lambda_{\rm irr}\approx 0.78$ for $R^\circ$, see \cref{tab:isospec1} below).

\subsection{Critical fixed point via Newton method}
\label{sec:Newton}

\begin{figure}
	\centering
	\includegraphics[width=0.9\textwidth]{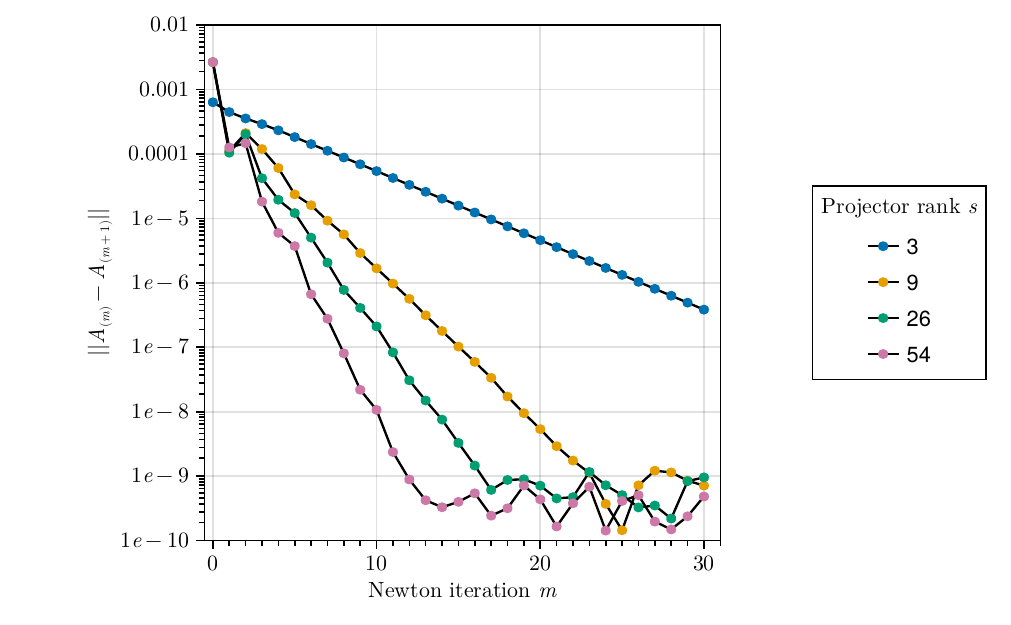}
	\caption{ 
		Convergence of the Newton method with a fixed approximate Jacobian for different ranks $s$ of the $P_s$ projector in \cref{eq:approxJ}. Figure from \cite{Ebel:2024nof}.
		\label{fig:newton_conv}
	}
\end{figure}

We next describe how the critical fixed point can be found via the Newton method. Unlike the "shooting" method, this will work only for $R^\circ$ but not for $R$ (see below), so we set everything up for $R^\circ$.

The Newton method used in \cite{Ebel:2024nof} solves the fixed point equation 
\beq
F(A):=A- R^\circ(A)=0
\eeq
via iterations
\beq
A_{(m+1)}= G(A_{(m)}),
\label{eq:newtonseq}
\eeq
where the Newton map $G$ is given by
\beq
G(A)=A- J^{-1} F(A)\,.
\label{eq:newton}
\eeq
For any invertible $J$ the fixed points of $G$ coincide with those of $R^\circ$. We will choose $J$ so that $G$ is a contraction in a neighborhood of the critical fixed point. Then the Newton iterates \eqref{eq:newtonseq} will converge exponentially fast for a sufficiently good initial approximation $A_{(0)}$. Namely, we will have:
\beq
\|A_{(m+1)}-A_{(m)}\|\sim \lambda^{m}
\label{eq:newtonconv}
\eeq
where $\lambda<1$ is the largest eigenvalue of $\nabla G$ at the fixed point.

The Newton method needs an RG map that is differentiable near the fixed point. In particular, this requires a robust gauge-fixing procedure. As long as the initial approximation is sufficiently good, the achievable accuracy is limited only by roundoff errors. This is unlike in the "shooting" method, where to improve the accuracy the initial tensor needs to be chosen ever closer to the critical manifold (reducing $\Delta t$ in Eq.~\eqref{eq:acc-shooting}). In principle, a suitable modification of the Newton method may converge to the fixed point when initialized from the nearest-neighbor Ising model tensor $A_{\rm NN}(1)$. The variant explored in \cite{Ebel:2024nof} had a smaller radius of convergence and relied on the "shooting" method to get a sufficiently good initial approximation.

We next describe the choice of $J$ in \eqref{eq:newton}. There is a hierarchy of Newton-like methods. For example, the exact Newton method would correspond to $J=\nabla F(A_{(m)})$, which would be rather expensive to compute and invert. Ref.~\cite{Ebel:2024nof} used instead the Newton method with a fixed approximate Jacobian which corresponds to choosing
	\beq
	J=I-P_s \nabla R^\circ(A_{(0)}) P_s\,,
	\label{eq:approxJ}
	\eeq 
	where $P_s$ is the orthogonal projector on the subspace spanned by the first $s$ $\mathbb{Z}_2$-even (see \cref{note:Z2even}) eigenvectors of $\nabla R^\circ(A_{(0)})$. For not too large $s$, these eigenvectors are affordable to compute using iterative Krylov subspace methods.

	\cref{fig:newton_conv} shows how this works in practice. Here $R^\circ$ is the rotating Gilt-TNR map with the same parameters as in \cref{fig:crit_convergence}. The initial approximation $A_{(0)}$, of accuracy $\delta \sim 3\times 10^{-3}$, was obtained applying the "shooting" method, determining the critical temperature to accuracy $\Delta t\sim 10^{-10}$ and performing 23 RG iterations of $R^\circ$. The subsequent iterations were performed using the Newton method. We see that the Newton iterations converge exponentially fast with an exponent which improves for larger $s$. This is natural because our construction \eqref{eq:approxJ} ensures that the first $s$ eigenvalues of $\nabla R^\circ$ around the fixed point are subtracted away, to the extent that one can ignore the difference between the Jacobian at $A_{(0)}$ and at the fixed point. See \cite[Sec.~IV~A]{Ebel:2024nof} for a detailed discussion. Eventually, the convergence stops and is replaced by a chaotic behavior where the distance between two consecutive iterates oscillates around $10^{-9}$, indicating roundoff errors. All tensors $A_{(m)}$ at the bottom of \cref{fig:crit_convergence} can be taken as excellent approximations of the rotating Gilt-TNR fixed point tensor $A_*^{\circ[30]}$, with accuracy $\delta\sim 10^{-9}$. For concreteness we will work below with the tensor $A_{(m_*)}$, $m_*=14$, from the $s=54$ Newton iteration sequence, which is the last iteration when $\|A_{(m+1)}-A_{(m)}\|$ decreases. 
	
	So we see that the Newton method works nicely for $R^\circ$. We will show next the Jacobian eigenvalues, and discuss why the Newton method could not be set up for $R$. 
	
	\subsection{CFT and Jacobian eigenvalues. Why rotation?}\label{sec:jacobian-eigenvalues}
	
	\begin{table}[ht]
		\centering
		\begin{tabular}{lccc}\toprule
			& $\mathbb{Z}_2$ & $\Delta$ & $ \ell$ \\
			\midrule
			$\mathds{1} $& + &0 & 0\\
			$T,\overline{T}$                & +              & 2        & $\pm 2$       \\
			$T\overline T$                  & +              & 4        & 0       \\
			$TT,\overline{T}\,\overline{T}$ & +              & 4        & $\pm 4 $      \\
			\midrule
			$\epsilon$                      & +              & 1        & $0$     \\
			$L_{-4}\epsilon+\ldots$         & +              & 5        & $4$     \\
			\midrule
			$\sigma$                        & $-$            & \sfrac 18    & 0       \\
			$L_{-3}\sigma+\ldots$           & $-$            & 3\,\sfrac 18     & 3       \\
			\bottomrule
		\end{tabular}
		\caption{\label{tab:2D}2D Ising CFT quasiprimary operators of scaling dimension $\Delta\le 5$, with their $\mathbb{Z}_2$ quantum number, scaling dimension $\Delta$, and spin $\ell=h-\bar{h}$ (see e.g.~\cite[Table 8.1]{DiFrancesco:1997nk}). Quasiprimary operators excluding the identity $\mathds{1}$ are referred to as ``nontrivial''.}
	\end{table}

	At long distances, the critical point of the 2D Ising model is described by a conformal field theory (CFT), which is the minimal model $\mathcal{M}_{3,4}$ ("2D Ising CFT"). One normally divides 2D CFT operators into Virasoro primaries and the rest which are Virasoro descendants. Here it will be useful to divide the operators a bit differently, namely into quasiprimaries and the rest which are $\text{SL}_2(\mathbb{C})$ descendants (derivatives). The 2D Ising CFT has three Virasoro primaries $\mathds{1},\sigma,\epsilon$ but infinitely many quasiprimaries. The spectrum of low-lying quasiprimaries is shown in \cref{tab:2D}.
	
	As explained in \cite{Ebel:2024nof}, when we diagonalize the Jacobian of an RG map around the fixed point, we will find eigenvalues of two kinds:
	\begin{itemize}
		\item
		Eigenvalues associated with the nontrivial quasiprimary operators $\mathcal{O}_{\Delta,\ell}$, whose values are given by
		\beq
		\lambda = \begin{cases} 
			b^{2-\Delta} & (\text{non-rotating RG map $R$}),\\
	    	 i^{\ell} b^{2-\Delta} & (\text{rotating RG map $R^\circ$}).
		\end{cases}
		\label{eq:eig-pred}
		\eeq
		\item
		Eigenvalues corresponding to derivatives of quasiprimary operators, whose values are non-universal and cannot be in general predicted from CFT.\footnote{An exception arises for special RG maps whose Jacobian coincides with the lattice dilatation operator, in which case the derivative eigenvalues become universal. This happens e.g.~for the ``frozen'' RG map in \cite{Lyu:2021qlw,PhysRevE.109.034111}, as noted in \cite{Ebel:2024nof} and discussed below in \cref{app:frozenJacs}.\label{ftnt:Jac_DSO}} 
		\end{itemize}

The first few eigenvalues of $\nabla R$ and $\nabla R^\circ$ at the approximate fixed points of the two maps are shown in \cref{tab:isospec1}, taken from \cite{Ebel:2024nof}. These results confirm that the quasiprimary eigenvalues are universal and given by \eqref{eq:eig-pred} and that there are also non-universal eigenvalues. It would be interesting to get concrete evidence for the intuition that the non-universal eigenvalues are indeed associated with the derivative operators \cite[Sec.~V]{Ebel:2024nof}. 
	
	\definecolor{light-blue}{HTML}{D1EDFF}
	\definecolor{light-red}{HTML}{FF999C}
	\begin{table}[]
		\centering
		\begin{NiceTabular}{c|cc|c|c}
			\CodeBefore
			\Body
			\toprule
			$\mathbb{Z}_2$ & $\mathcal{O}$ & $\lambda_{\rm CFT}=2^{2-\Delta_\mathcal{O}}$ & $\lambda$ of $\nabla R$ &  $\lambda$ of $\nabla R^{\circ}$ \\
			\midrule
			$+$            & $\epsilon$            & 2                                            & 1.9996                  & 1.9996                          \\
			$+$            & $T$                   & 1                                            & 1.0015                  &  $-1.0010  $                       \\
			$+$            & $\overline{T}$             & 1                                            & 0.9980                  & $ -0.9982 $                        \\
			$+$            &                       &                                              & 0.6322                  &  0.7757                          \\
			$+$            &                       &                                              & $0.5941\pm 0.1195 i$    &  0.6209                          \\
			&                       &                                              & \ldots                  &  \ldots                          \\
			\midrule
			$-$            & $\sigma$              & $2^{15/8}\approx 3.668016$                   & 3.6684                  &  3.668014                        \\
			$-$            &                       &                                              & 1.5328                  &  $0.0027\pm 1.5364i$             \\
			$-$            &                       &                                              & 1.5300                  &  0.8600                          \\
			$-$            &                       &                                              & 0.8869                  &  $0.6318 \pm 0.0583i$            \\
			&                       &                                              & \ldots                  &  \ldots                      
			\\\bottomrule   
		\end{NiceTabular}
		\caption{Column 1: $\mathbb{Z}_2$ quantum number. Column 2,3 (rows with entries in those columns): the first few low-dimension CFT quasiprimaries, and their exact eigenvalues. Columns 4,5: The first few largest, in absolute value, eigenvalues of the Jacobians $\nabla R$ and $\nabla R^\circ$ at the approximate fixed points of the two maps. Rows without entries in columns 2,3 show RG eigenvalues corresponding to derivative interactions, which are not universal. Table from \cite{Ebel:2024nof}.}
		\label{tab:isospec1}
	\end{table}
	
	Looking at \cref{tab:isospec1}, we see that $\nabla R$ has two eigenvalues 1 at the fixed point, corresponding to spin-2 quasiprimaries $T,\bar{T}$, components of the CFT stress tensor operator. Eigenvalues 1 are problematic for the Newton method because the Jacobian $\nabla F=I-\nabla R$ would not be invertible at the fixed point. It is to solve this issue that Ref.~\cite{Ebel:2024nof} introduced the rotating RG map $R^\circ$. Rotation turns the stress tensor eigenvalues from 1 to $-1$, see \eqref{eq:eig-pred} and \cref{tab:isospec1}. As a result, the Jacobian $\nabla F=I-\nabla R^\circ$ is perfectly invertible, and the Newton method can be set up. 
	
	Relative errors for the scaling dimensions of $\sigma,\epsilon, T,\overline T$ extracted from the corresponding $\nabla\mathcal{R}$ and $\nabla\mathcal{R}^\circ$ eigenvalues are summarized in \cref{tab:Jac-errors}. The errors are similar for the two RG transformations, except for the uncanny accuracy that $\nabla\mathcal{R}^\circ$ gives for $\sigma$.
	 
	\begin{table}[]
		\centering
		\begin{tabular}{lccc}
			\toprule
			& & $\nabla\mathcal{R}$ & $\nabla\mathcal{R}^\circ$ \\
			\midrule
			$\Delta=\text{\sfrac{1}{8}}$ & $\sigma$ & 0.13\%              & 0.00066\%                 \\
			$\Delta=1$ & $\epsilon$   & 0.03\%              & 0.026\%                   \\
			$\Delta=2$ & $T,\overline T$   & 0.15\%              & 0.13\%                   \\\bottomrule
		\end{tabular}
		\caption{Relative errors for scaling dimensions of quasiprimaries of $\Delta \le 2$ extracted from $\nabla\mathcal{R}$ and $\nabla\mathcal{R}^\circ$ eigenvalues.}
		\label{tab:Jac-errors}
	\end{table}
	
	This finishes our brief review of the main results of \cite{Ebel:2024nof}.

\section{Transfer matrix}
	\label{sec:transfer}

Apart from the Jacobian eigenvalues discussed above, there are two other ways of extracting CFT scaling dimensions from tensor RG: transfer matrix (TM) \cite{TEFR} and lattice dilatation operator \cite{TNRScaleTrans}. In this section we will consider the transfer matrix method. We first discuss the theory behind this method, and then present the numerical results.

\subsection{Theory}
\label{sec:transfer-theory}

The use of TM on a finite chain to extract CFT eigenvalues goes back to \cite{PhysRevLett.56.742}. For a generic critical tensor, one would have to increase the TM width to improve the accuracy. Here instead we will be building the TM from a fixed-point tensor. In that case a TM of constant width is sufficient, as was first pointed out in \cite{TEFR}. Of course, one still has to increase the bond dimension to reduce the truncation errors.

Let $h,v$ be two integers. We will need to consider tensor network partition functions with twisted boundary conditions. We denote by $Z_{h,v}(A,N\times M)$ the partition function of $N\times M$ tensor network built out of tensor $A$ such that the bottom legs  $i$ are contracted with the top legs $i+h$ modulo $N$, and left legs $j$ are contracted with right legs $j+v$ modulo $M$:
\beq
\myinclude[scale=0.8]{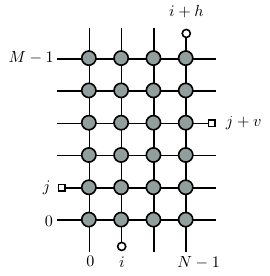}.
\label{eq:gluing}
\eeq 
When only one of two numbers $h,v$ is nonzero, the contraction procedure indicated in \eqref{eq:gluing} gives a tensor network which is locally indistinguishable from the periodic boundary conditions case $h=v=0$, i.e.~it has no defects. Only such cases will be considered below. (When both $h\ \text{mod}\ N$ and $v\ \text{mod}\ M$ are nonzero, this introduces a defect into the network.)

\begin{definition}\label{def:friendly}Let $\mathcal{R}$ be a non-rotating RG map with scale factor $b$, and let $h,v$ be two nonnegative integers divisible by $b$, exactly one of which is nonzero. We will say that $\mathcal{R}$ is $(h,v)$-twist friendly, if it preserves the twisted partition function up to reducing all dimensions and twists by the same factor $b$. I.e., for $N,M$ divisible by $b$, we have:
\beq
Z_{h,v}(A,N\times M)=Z_{h/b,v/b}(\mathcal{R}(A),N/b\times M/b)\,.
\label{eq:nonrotZhv}
\eeq
We will also say that a rotating RG map $\mathcal{R}^\circ = \Gamma_{\pi/2}\mathcal{R}$ is $(h,v)$-twist friendly if this holds for the non-rotating RG map $\mathcal{R}$.
\end{definition}

The twist friendliness condition is needed for the validity of some TM spectrum properties stated below. All RG maps known to us, and in particular Gilt-TNR, are $(h,0)$- and $(0,v)$-twist friendly for any $h,v$ divisible by $b$. Intuitively, twist friendliness is a consequence of ``locality'' of the RG map, and it appears to have been tacitly assumed in \cite{TEFR}. Rather than trying to formalize the connection to locality, we prefer to state this condition explicitly.

Let us proceed to the construction of the TMs. To avoid some trivial normalization factors in the spectrum, it's most convenient to construct the TMs not from the fixed point tensor $A_*$ but from its particular rescaling $A^*$ which we now define.

Recall that the fixed point tensor $A_*$, which has unit norm, satisfies equation $R(A_*)=A_*$ where $R$ is the normalized RG map (which may be rotating or non-rotating). Acting on the same tensor with the non-normalized map $\mathcal{R}$ we will get $\mathcal{R}(A_*) = \mathcal{N} A_*$ where $\mathcal{N}>0$ is a constant. The non-normalized RG map is typically homogeneous with degree $p=b^2$, namely $\mathcal{R}(\alpha A)=\alpha^p A$ for any $A$ and any $\alpha>0$. E.g.~Gilt-TNR has $p=4$. We define $A^*=\mathcal{N}^{-1/(p-1)} A_*$. We have $\mathcal{R}(A^*)= A^*$, i.e.~it's a fixed point of $\mathcal{R}$. 

The TM will be constructed from $r\ge 1$ copies of the tensor $A^*$. For an exact fixed point tensor, $r=1$ would already be enough to extract the exact CFT spectrum. Working with a numerical approximate fixed point, using $r>1$ improves the accuracy of the method.

We will consider two kinds of TMs, ``direct'' and ``crossed''. The direct TM $\calM_r$ is defined by the following equation:
\beq
\myinclude[scale=0.8]{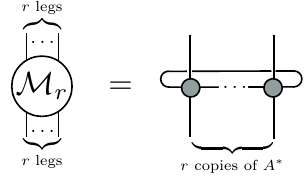}\qquad(\text{direct}),
\eeq 
Its eigenvalues are given in terms of the scaling dimensions $\Delta_k$ of CFT local operators (both quasiprimaries and descendants), by\footnote{To avoid confusion, we will denote Jacobian eigenvalues $\lambda_k$, TM eigenvalues $\mu_k$, and lattice dilatation operator eigenvalues $\rho_k$.}
\beq
\mu_k  = e^{-\frac{2\pi}{r} (\Delta_k-c/12)}\qquad (\text{direct}).\label{eq:TMeigs1}
\eeq
The direct TM commutes with the translation generator $\calT$ which permutes the legs circularly:
\beq
\myinclude[scale=0.8]{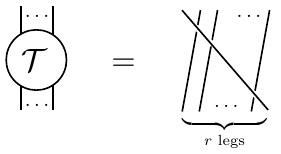}.
\label{eq:transl}
\eeq
Since $\calT^r=I$, the eigenvalues of $\calT$ are $r$-th roots of unity. Let us assume that the RG map is $(h,0)$-twist friendly for $b|h$. Then more can be said, namely that the eigenvalues of $\calT$
are given by 
\beq
\tau_k = e^{i \frac{2\pi}{r}\ell_k}\,
\label{eq:Teigs}
\eeq
in terms of the spins $\ell_k$ of the CFT operators. This fact can be used to determine $\ell_k\text{ mod } r$.

Finally we define the crossed TM $\widetilde{\calM}_r$ by
\beq
\myinclude[scale=0.8]{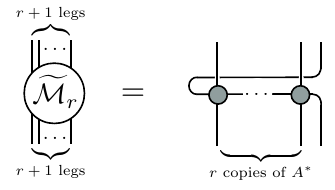}\qquad(\text{crossed})\,.
\eeq
Its eigenvalues are given by
\beq
		\mu_k = e^{-\frac{2\pi r}{1+r^2} (\Delta_k-c/12) + i\frac{2\pi}{1+r^2}\ell_k}\qquad(\text{crossed})\,.
		\label{eq:TMeigs2}
\eeq
This equation holds provided that the RG map is $(0,v)$-twist friendly for $b|v$. We see that the phase of crossed TM eigenvalues determines the spin $\ell_k$ modulo $1+r^2$. 

Most of the literature so far used the direct $\calM_r$ with $r=1$ or $r=2$ (but without using the information from $\calT$ eigenvalues) \cite{Evenbly-Vidal,LoopTNR,Hauru:2015abi,TNRScaleTrans,Bal:2017mht,GILT,IinoMoritoKawashima,Lyu:2021qlw,PhysRevE.109.034111,2023arXiv230617479H}. The possibility to use the crossed $\widetilde{\calM}_r$ (only for $r=1$) was mentioned in \cite{TEFR} without detailed justification, and to our knowledge it has never been tested. 

Here we will give a brief reminder why Eq.~\eqref{eq:TMeigs1} holds, for $r=1$, following \cite{TEFR}. See \cref{app:TM} for general $r$, and for the translation generator and the crossed TM eigenvalues.

Consider an $N\times M$ tensor network built out of the critical fixed point tensor $A^*$. For $N,M\gg 1$, we expect that the tensor network partition function built out of any tensor on the critical manifold, in particular out of the fixed point tensor $A^*$, equals the CFT partition function up to an extra factor, corresponding to a volume-proportional term in the free energy:
\beq
Z(A^*, N\times M)\approx e^{-fNM} Z_{\rm CFT}(N\times M)\qquad(N,M\gg 1).
\label{eq:ZZCFT}\eeq
The CFT partition function equals \cite{DiFrancesco:1997nk}
\beq
Z_{\rm CFT}= \sum_{k=0}^\infty e^{-\frac{M}{N} {2\pi}(\Delta_k-c/12)},
\label{eq:ZCFT}
\eeq
while the tensor network partition function can be reduced, applying the RG transformation several times, to the partition function of size $1\times \frac{M}{N}$, which in turn can be expressed via the eigenvalues of the direct TM $\calM_r$, $r=1$ as:
\beq
Z(A^*, N\times M) = Z(A^*,1\times \frac{M}{N}) = \sum_{k=0}^\infty (\mu_k )^{\frac MN}\,.
\label{eq:ZTMtocomp}
\eeq
(This only makes sense if $N$ divides $M$, and in the precise argument in \cref{app:TM} we will only consider this case.)
Since \eqref{eq:ZZCFT} needs to hold for $N,M\gg 1$, we conclude first of all that the nonuniversal (i.e.~not predicted by CFT) part of the free energy $f$ vanishes, and second that $\mu_k$ should be given by the $r=1$ case of \eqref{eq:TMeigs1}.\footnote{As a byproduct of this argument, we see that Eq.~\eqref{eq:ZZCFT} holds exactly for any $N$ a power of $b$, $M$ an integer multiple of $N$, and not just for $M,N\gg 1$.}

A note on the computational cost is in order. Clearly, working with the finite bond dimension, we cannot get the full CFT spectrum right. So our target is typically the first $k$ scaling dimensions, where $k$ is some number much smaller than the full dimension $K$ of space in which the TM acts. We have $K=\chi^r$ for the direct TM $\mathcal{M}_r$ and $K=\chi^{r+1}$ for the crossed TM $\widetilde{\mathcal{M}}_{r}$. To get the first $k$ eigenvalues, we use Arnoldi's method that scales as $O(k \times T)$, where $T$ is the "matrix application time", i.e.~the time required to act with the TM on a vector. If we have the TM stored as a $K\times K$ matrix, then $T=K^2$. But computing the matrix has its own cost in time (as well as in memory). For $r\ge 2$, the case of interest below, the faster option is not to compute the full TM but to evaluate its action "on the fly", given its expression in terms of $A_*$. This gives the following optimal application times $T$:\footnote{We thank the anonymous referee for asking us to identify this optimal scaling. Our original code, used to generate Table \ref{tab:TMerrors}, computed the full matrix and was consequently slower, scaling for $r=2$ as $\chi^6$ (direct TM) and $\chi^7$ (crossed TM). To obtain the results in \cref{tab:TMerrorsr4}, the code was updated; the new version \cite{our-code} now includes scripts for computing the TM spectra for $r=4$ with the optimal scaling presented in \cref{eq:TMscaling}. }
	\begin{align}
		&\text{direct TM $\mathcal{M}_r$:}\quad \sim\chi^5\ (r=2),\quad \sim \chi^{r+4}\ (r\ge 3)\,,\label{eq:TMscaling}\\ 
		&\text{crossed TM $\widetilde{\mathcal{M}}_r$:} \quad \sim\chi^{r+3}\ (r\ge 2)\,.\nonumber
	\end{align}

\subsection{Numerical results}
		\label{sec:transfer-numerics}
		
\begin{figure}
		\centering
		\includegraphics[width=\textwidth]{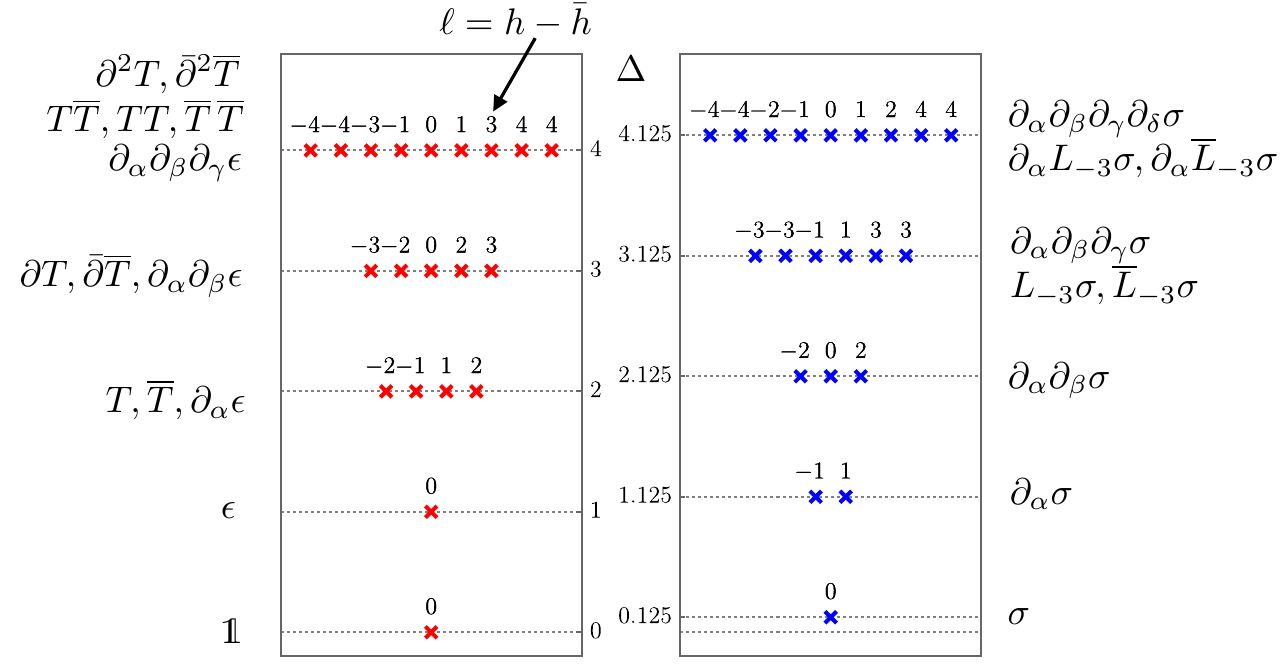}
		\caption{
			\label{fig:CFTspec} Exact spectrum of the critical 2D Ising CFT. 
		}
\end{figure}

In \cref{fig:CFTspec}, we show the exact spectrum of local operators of the critical 2D Ising CFT up to scaling dimension $\Delta=4\oneeighth$. This spectrum consists of Virasoro descendants of $\mathds{1}$ and $\epsilon$ ($\mathbb{Z}_2$-even sector, left), and of $\sigma$ ($\mathbb{Z}_2$-odd sector, right). The number near each cross shows the local operator spin.

These exact results will be now compared with TM results. We will present the results for the direct and crossed TM's for $r=2$ and $r=4$.\footnote{ We pushed the computation to $r=4$ for a fair comparison with LDO, which has an application time scaling as $\chi^8$. The crossed TM with $r=4$ (application time scaling as $\chi^7$) is faster than LDO, and a fairer comparison would involve the $r=5$ crossed TM. However, the $r=5$ computation proved too memory-demanding. The memory requirements could be improved by leg compression, as in the LDO case (see \cref{app:compression}), but this was not attempted as the $r=4$ crossed TM already yielded excellent results. We thank the anonymous referee for asking us to explore wider TMs. \label{foot:r45}} We also have two approximated fixed point tensors - for the non-rotating map $\mathcal{R}$ and for the rotating map $\mathcal{R}^\circ$. The spectra of the Jacobian for these two approximate fixed points were shown in \cref{tab:isospec1}. We thus have eight sets of results (four TMs and two approximate fixed points). 

In each of these eight cases, we compute TM eigenvalues and we extract the scaling dimensions using the formulas from the previous section. We extract the central charge $c$ using the fact that the unit operator must have scaling dimension 0. In the direct TM case, we compute spin modulo $r=2,4$ using the translation generator eigenvalues. In the crossed TM case, we also compute spin modulo $1+r^2=5,17$ from the phase of the TM eigenvalues.

We start with the results for $r=2$. We first describe the qualitative results, and then show some numbers.{ In each of the four $r=2$ cases, the central charge and scaling dimensions, the number of states, and their spins extracted from TM eigenvalues are in excellent agreement with the CFT up to $\Delta=3$ ($3\oneeighth$) in the even (odd) sector. In particular, up to this level the scaling dimensions are clearly approximately quantized in agreement with the CFT. Starting from $\Delta=4$ (4\sfrac 18) in the even (odd) sector, the TM spectrum ceases to be clearly quantized and TM levels cannot be clearly assigned to CFT levels. So we do not show the results beyond $\Delta=3\oneeighth$.

The numerical results for the TM spectrum can be found in a spreadsheet accompanying this article. The accuracy of the central charge prediction and of the scaling dimensions up to $\Delta=3\oneeighth$ is summarized in \cref{tab:TMerrors}. These errors can be compared with the errors for the scaling dimensions of quasiprimaries determined from the eigenvalues from the RG map Jacobians, \cref{tab:Jac-errors}.

The errors displayed in \cref{tab:TMerrors} for $\mathcal{R}$ and $\mathcal{R}^{\circ}$ are comparable. We do not see a significant improvement of scaling dimensions in the $\mathcal{R}^{\circ}$ case, even though for these results we used a better fixed point approximation $A_{(m_*)}$ found by the Newton method. In fact, some scaling dimensions in the $\mathcal{R}^{\circ}$ case are even slightly worse than in the $\mathcal{R}$ case. Our Newton method for the rotating algorithm began with a tensor $A_{(0)}$ which is found using the shooting method with $\Delta t \sim 10^{-10}$ and whose accuracy $\delta \sim 3 \times 10^{-3}$ is rather poor. We additionally computed the scaling dimensions using TM for this less accurate fixed point approximation $A_{(0)}$. The result can be found in the spreadsheet accompanying this article. A comparison of TM results for $A_{(0)}$ and $A_{(m_*)}$ showed that the scaling dimensions precision for $A_{(m_*)}$ is only slightly better for scaling dimensions in \cref{tab:TMerrors} in the direct TM case. In the crossed TM case, some scaling dimensions for $A_{(m_*)}$ have slightly larger errors than those for $A_{(0)}$, but they are still comparable in general.

\begin{table}[]
	\centering
	\begin{tabular}{lllll}
		\toprule
		& $\mathcal{R}$, direct & $\mathcal{R}^\circ$, direct & $\mathcal{R}$, crossed & $\mathcal{R}^\circ$, crossed \\
		\midrule
		$c$            & 0.1\%                  & 0.05\%                       & 0.01\%                  & 0.014\%                       \\\midrule
		$\Delta=\oneeighth$ & 0.06\%                 & 0.01\%                       & 0.003\%                 & 0.031\%                       \\
		$\Delta=1$     & 0.19\%                 & 0.20\%                       & 0.08\%                  & 0.09\%                        \\
		$\Delta=1\,\oneeighth$ & 0.29\%                 & 0.28\%                       & 0.08\%                  & 0.04\%                        \\
		$\Delta=2$     & 0.76\%                 & 0.66\%                       & 0.18\%                  & 0.09\%                        \\
		$\Delta=2\,\oneeighth$ & 1.71\%                 & 1.14\%                       & 0.30\%                  & 0.18\%                        \\
		$\Delta=3$     & 1.24\%                 & 1.09\%                       & 0.30\%                  & 0.50\%                        \\
		$\Delta=3\,\oneeighth$ & 1.46\%                 & 1.80\%                       & 0.44\%                  & 1.89\%       \\                
		\bottomrule
	\end{tabular}
	\caption{Relative errors for the central charge and scaling dimensions extracted via direct and crossed TM, with $r=2$, applied to the approximate fixed points of $\mathcal{R}$ and $\mathcal{R}^\circ$. For degenerate scaling dimensions, we show the maximal value of the error. In all cases shown in this table, the number of states and the partially available spin information agree between the TM and the CFT. }
	\label{tab:TMerrors}
	\end{table}

As mentioned, the number of states and the partially available spin information agrees between the TM and the CFT for all cases shown in \cref{tab:TMerrors}. Let us give a couple of examples. CFT predicts 6 $\mathbb{Z}_2$-odd local operators of $\Delta=3\oneeighth$, of spins 
\beq
\ell = \pm 3,\pm1,\pm 3.\label{eq:exact-spin}
\eeq
The $r=2$ direct TM applied to the approximate fixed point of $\mathcal{R}^\circ$ has exactly 6 eigenvalues in the range of interest, of scaling dimensions 
\beq
3.0689, 
3.0791, 
3.1392, 
3.1397, 
3.1507,
3.1591\qquad(\text{maximal deviation 1.80\%})
\eeq
and their spin mod 2 extracted from the $\mathcal{T}$ eigenvalue are $1,1,1,1,1,1$, as it should be.
 
Similarly, the $r=2$ crossed TM applied to the approximate fixed point of $\mathcal{R}^\circ$ has 6 eigenvalues of scaling dimensions 
\beq
3.1251,
3.1251,
3.1338,
3.1338,
3.1841,
3.1841\qquad(\text{maximal deviation 1.89\%}).
\eeq
Their spin mod 5 extracted from the phase of the eigenvalues are 
\beq
\pm 1.96
\pm 1.03,
\pm 1.80,
\label{eq:TM-spin}
\eeq
rather close to the exact values \eqref{eq:exact-spin} mod 5, which would be $\pm 2,\pm 1,\pm 2$. 

\begin{table}[]
	\centering
	\begin{tabular}{lllll}
		\toprule
		& $\mathcal{R}$, direct & $\mathcal{R}^\circ$, direct & $\mathcal{R}$, crossed & $\mathcal{R}^\circ$, crossed \\
		\midrule
		$c$            & 0.05\%                  & 0.005\%                       & 0.04\%                  & 0.012\%                       \\\midrule
		$\Delta=\oneeighth$ & 0.04\%                 & 0.03\%                       & 0.03\%                 & 0.031\%                       \\
		$\Delta=1$     & 0.02\%                 & 0.05\%                       & 0.01\%                  & 0.04\%                        \\
		$\Delta=1\,\oneeighth$ & 0.04\%                 & 0.07\%                       & 0.01\%                  & 0.04\%                        \\
		$\Delta=2$     & 0.16\%                 & 0.17\%                       & 0.09\%                  & 0.11\%                        \\
		$\Delta=2\,\oneeighth$ & 0.29\%                 & 0.27\%                       & 0.14\%                  & 0.15\%                        \\
		$\Delta=3$     & 0.46\%                 & 0.40\%                       & 0.25\%                  & 0.26\%                        \\
		$\Delta=3\,\oneeighth$ & 0.71\%                 & 0.54\%                       & 0.38\%                  & 0.36\%       \\                
		$\Delta=4$ & 0.88\%                 & 0.62\%                       & 0.54\%                  & 0.51\%       \\                
		$\Delta=4\,\oneeighth$ & 1.75\%                 & 1.08\%                       & 0.82\%                  & 0.72\%       \\                
		\bottomrule
	\end{tabular}
	\caption{Relative errors for the central charge and scaling dimensions extracted via direct and crossed TM, with $r=4$, applied to the approximate fixed points of $\mathcal{R}$ and $\mathcal{R}^\circ$. For degenerate scaling dimensions, we show the maximal value of the error. In all cases shown in this table, the number of states and the partially available spin information agree between the TM and the CFT.}
	\label{tab:TMerrorsr4}
\end{table}

Let us now discuss the $r=4$ results. \cref{tab:TMerrorsr4} displays the errors in scaling dimensions for all four cases (crossed/direct TM and rotating/non-rotating algorithm fixed points). Aside from a few outliers, there is an overall improvement in precision compared to the $r=2$ case. The improvement is especially prominent in the high part of the spectrum, where the $r=4$ TM allows for the inspection of more states. As with the $r=2$ case, the results for $\mathcal{R}$ and $\mathcal{R}^\circ$ have comparable errors, with neither showing a distinct advantage.  

Let us provide some rationale for why increasing $r$ should decrease the numerical error, as we observe. 
The derivation of our formulas for the scaling dimensions from TM eigenvalues \cref{eq:TMeigs1,eq:TMeigs2} is based on matching partition functions on a long torus (see \cref{app:TM}). These eigenvalues decay exponentially with $\Delta$, as
\beq
\label{eq:muDelta}
\mu\sim e^{-\frac{2\pi}{r} \Delta} (\text{direct});\quad e^{-\frac{2\pi r}{1+r^2} \Delta} (\text{crossed})\,.
\eeq
Increasing $r$, the exponential decrease of the eigenvalues gets slower, and so sensitivity to higher eigenvalues gets enhanced. For example, for the crossed TM we pass from $\approx 0.08^{\Delta}$ for $r=2$ to $\approx 0.2^{\Delta}$ for $r=4$. This is a likely reason which explains the improved numerical resolution of wider TM's in the higher part of the spectrum.


For all cases in \cref{tab:TMerrorsr4}, the number of states and the partially available spin information agree between the TM and the CFT. In particular, the $r=4$ crossed TM allows for resolving all spins modulo 17, which is sufficient to assign exact spin values for all computed states, since the spin on levels $n$ and $n+1/8$ in the 2D Ising CFT varies from $-n$ to $n$. For example, at level $\Delta = 4 \oneeighth$ the crossed TM for $\mathcal{R}^\circ$ gives
\beq
\begin{matrix}
	\Delta= & 4.1332 \ \  & 4.1332 \ \ & 4.1360 \ \ & 4.1360 \ \ & 4.1368 \ \ & 4.1403 \ \ & 4.1403  \ \ & 4.1546 \ \ & 4.1546  \\
	\ell= & 2.02 & -2.02 & 4.05 & -4.05 & 0.00 & 2.05 & -2.05 & 4.08 &  -4.08
\end{matrix}
\eeq

To see where the $r=4$ TM method breaks down, we pushed the crossed TM computation to scaling dimensions higher than those reported in \cref{tab:TMerrorsr4} (only for the $\mathcal{R}^{\circ}$ fixed point). We found approximate level quantization, with the number of states in agreement with the CFT, at levels up to $\Delta=6 \oneeighth$. Since the exact levels in each $\mathbb{Z}_2$ sector are separated by 1, a reasonable criterion for the breakdown of approximate level quantization is an absolute error exceeding $0.5$. This first occurs at 7th state on the $\Delta=7$ level. In total, the 116 states before this breakdown are determined with a maximum absolute (relative) error of 0.12 (2\%). 

Finally, we compared the $r=4$ TM results for $\mathcal{R}^\circ$ using both $A_{(0)}$ and $A_{(m_*)}$. A consistent improvement was observed for most scaling dimensions when using the more precise critical point $A_{(m_*)}$. The exception was the smallest scaling dimension, $\Delta=\oneeighth$, for which the error in the crossed TM case is slightly bigger for $A_{(m_*)}$. Here is the crossed TM example:
\beq \label{eq:r4impr}
\begin{matrix}
	\Delta & \oneeighth & 1 & 1\oneeighth & 2 & 2 \oneeighth & 3 & 3\oneeighth & 4 & 4 \oneeighth \\
	A_{(0)} & 0.028\% & 0.07\% & 0.07\% & 0.14\% & 0.19\% & 0.29\% & 0.40\% & 0.54\% & 0.76\% \\
	A_{(m_*)}  & 0.031\% & 0.04\% & 0.04 \% & 0.11\% & 0.15\% & 0.26\% & 0.36\% & 0.51\% & 0.71\%  \\  
\end{matrix}
\eeq
Here, the first row lists the exact scaling dimensions, while the second and third rows list the maximal errors for the scaling dimensions obtained using $A_{(0)}$ and $A_{(m_*)}$, respectively. The third row corresponds to the last column of \cref{tab:TMerrorsr4}. The full data set is available in the accompanying spreadsheet. The data in \cref{eq:r4impr} shows that using the more precise fixed point $A_{(m_*)}$ yields a noticeable improvement for low-lying states (with the exception of $\Delta=\oneeighth$), an effect that diminishes for larger scaling dimensions.

Overall, we saw that the difference in errors between the computations using $A_{(0)}$ and $A_{(m_*)}$ is not dramatic and consistent improvements appear for TM with $r=4$ but not $r=2$. This suggests that the error is dominated by truncation effects rather than by the error of the fixed point approximation, at least for a bond dimension of $\chi=30$. It is possible that for larger bond dimensions, the precision of the fixed point tensor approximation will play a more significant role.

\section{Lattice dilatation operator}\label{sec:DSO}
		 
In this section, we discuss the lattice dilatation operator (LDO) method for extracting CFT scaling dimensions. First proposed in \cite{TNRScaleTrans}, this technique is less frequently used than the TM method (the list of known to us references includes \cite{TNRScaleTrans, Evenbly-TNR-website, Evenbly-review, Bal:2017mht,IinoMoritoKawashima}). Here, we first give the theoretical background, then discuss the LDO for the Gilt-TNR algorithm, and finally present our numerical results.

\subsection{LDO definition}\label{sec:DSO-theory}

Let $\mathcal{R}$ be the non-normalized, non-rotating RG map for which $A^*$ is the critical fixed point: $\mathcal{R}(A^*)=A^*$.\footnote{See the discussion after Definition \ref{def:friendly}.} We start by taking a network built out of $A^*$, with periodic boundary conditions, and introducing a $2 \times 2$ hole in it by removing a $2 \times 2$ block of tensors.\footnote{One could imagine a generalization where an $r'\times r'$ hole is removed, for $r'>2$. This will be somewhat analogous to the TM method with $r=4r'$, and one may expect improved accuracy for higher $r'$. However the computation cost grows quickly with $r'$ and in this paper we will only consider $r'=2$,  the minimal value for which we found a realization of this idea within Gilt-TNR.} The result is no longer a partition function which is a single number, but a vector indexed by possible assignments of indices to the noncontracted bonds around the hole. We will refer to it as a {\bf network with a $2 \times 2$ hole} and denote it as a vector as follows:\footnote{The upcoming discussion generalizes straightforwardly to the case of holes of arbitrary size and shape. Additionally, one may consider networks away from criticality as it was done in Ref.~\cite{TNRScaleTrans}.}
\begin{equation}\label{eq:SO-theory1}
	\myinclude{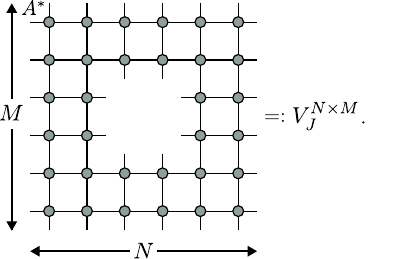} 
	\eeq
Here $J$ represents the collections of indices of noncontracted bonds around the $2\times 2$ hole.

Now, let us define an LDO. Let us think what happens when we try to apply $\mathcal{R}$ to a network with a hole. Most RG maps, including Gilt-TNR, act locally, and we will have no problem applying it away from the hole. So away from the hole we will recover again a tensor network made of $A^*$, rarefied by factor $b$. However, close to the hole, the action of the RG map will be inevitably disturbed by its presence. Let us suppose that our RG map can be performed in the presence of a hole, accounting for this presence by a linear operator $\calD=\calD^I{}_J$, namely: 
\begin{equation}
	\label{eq:SO-theory2}
	V^{N \times M}_J= V^{N/b \times M/b}_I \mathcal{D}^I{}_J\qquad \forall J\,.
\end{equation} 
Graphically the same equation is represented as follows:
\begin{equation}\label{eq:SO-theory3}
	\myinclude{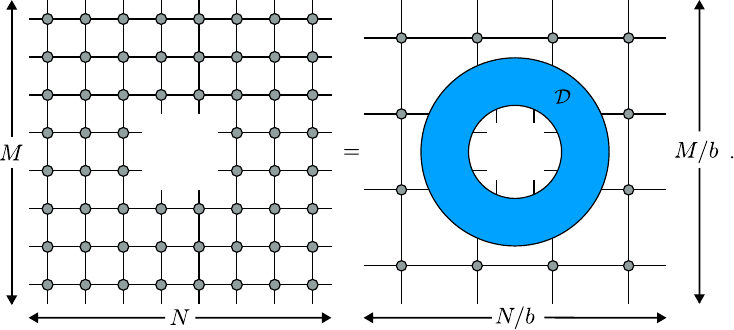}
\end{equation}
If this holds, we call $\mathcal{D}$ an LDO. We may think that it is composed of various pieces of the RG map, which cannot be recombined due to the presence of the hole, and are left behind to form $\mathcal{D}$.

Some remarks are in order here. First, note that we posit the existence of an operator $\mathcal{D}$ which acts from the networks with $2\times2$ holes to networks with $2\times 2$ holes. It's not obvious that the hole shape may be so preserved. Instead, it can happen that the hole shrinks or expands in size. The shrinking hole case will be discussed in \cref{app:frozenJacs}, where we will explain some aspects of the results of \cite{Lyu:2021qlw,PhysRevE.109.034111} (see \cref{ftnt:Jac_DSO}). Here in the main text we focus on the $2\times2$ hole preserving case since this turns out to be relevant for Gilt-TNR.

Second, from the logic by which we were led to \eqref{eq:SO-theory2}, appealing to locality of the RG map, it is plausible that if we create multiple $2\times 2$ holes in the network, then to obtain the same state after an RG step, we should simply surround each hole with the $\mathcal{D}$ operator as shown in \cref{eq:SO-theory3}. We will assume that this is the case. Again one can check that this holds for Gilt-TNR.

For a rotating RG map $\mathcal{R}^{\circ}$ given by \eqref{eq:Rwithrot}, \cref{eq:SO-theory2,eq:SO-theory3} are modified accordingly, swapping $M/b$ and $N/b$ in the r.h.s. The analogue of \eqref{eq:SO-theory2} takes the form
\begin{equation}
	\label{eq:SO-theory-rot}
	V^{N \times M}_J= V^{M/b \times N/b}_I \mathcal{D^\circ}^I{}_J\qquad \forall J\,.
\end{equation} 
The rotating LDO is related to the LDO for the corresponding non-rotating map by
\beq
\mathcal{D^\circ}^I{}_{J} = (\Gamma_{2\times 2})^I{}_{I'} \mathcal{D}^{I'}{}_J 
\label{eq:LDOcirc}
\eeq
where $\Gamma_{2\times 2}$ permutes the external $(I)$ LDO indices in a way corresponding to the $\pi/2$ rotation of the $2\times2$ hole (it is the analogue of $\Gamma_{\pi/2}$ appearing in \eqref{eq:Rwithrot} but for a tensor with 8 legs). Note that the internal $(J)$ indices are left intact.

\subsection{LDO eigenvalues and scaling operators}
\label{sec:LDOeigs}

The most amazing property of LDO is that it allows us to construct local scaling operators. These are local defects inserted into the tensor networks whose correlation functions have exact discrete scale invariance.\footnote{For other nonperturbative approaches to RG, the construction of local scaling operators is usually nontrivial. See e.g.~\cite{Giuliani:2024rwj} for some recent work in this direction, which however constructed only ``almost'' scaling operators.} Namely, let $O$ be an eight-legged tensor that is an eigenvector of $\mathcal{D}$ with the eigenvalue $\rho$, i.e.
\beq
\mathcal{D}^I{}_J\, O^J = \rho O^I
\eeq
or graphically
\beq
\myinclude{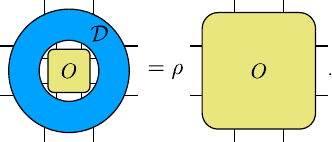}
\eeq
We will now show two facts:
\begin{itemize}
	\item
	Each LDO eigenvector $O$ is a local scaling operator.
	\item
	The LDO eigenvalues are related to CFT scaling dimensions and spins by:
	\begin{equation}\label{eq:SO-theory4}
		\rho=\begin{cases}
			b^{-\Delta} & (\text{non-rotating RG map }\mathcal{R}),\\
			i^{\ell}b^{-\Delta} & (\text{rotating RG map }\mathcal{R}^{\circ}).
		\end{cases}
	\end{equation}
	\end{itemize}
	
Consider a two-point correlator defined by inserting two copies of $O$ into the tensor network made of $A^*$:
\begin{equation}\label{eq:DSO5}
 \myinclude{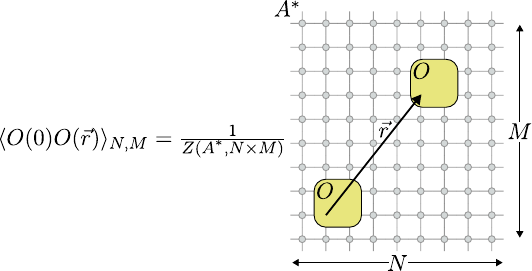}\, .
\end{equation}
In other words, $\<O(0)O(\vec{r})\>_{N, M}$ is, up to normalization, the contraction of the network with two $2 \times 2$ holes at 0 and $\vec{r}$ with two copies of $O$. We assume that $\vec{r}$ is large enough so that the holes are well separated. Also, $\vec r$ should be divisible by $b$.

Let us perform an RG step and see what happens with the correlator \eqref{eq:DSO5}. We assume first that the RG map is non-rotating (the rotating case is similar and it will be considered later). First of all, the RG map rescales the distance between the $O$ tensors by the factor of $1/b$. Second, we may work out what happens around the holes (filled by the $O$ tensors) using the definition of the LDO (recall we are assuming that LDO can also be used for multiple holes). Both $O$ tensors end up being surrounded by the LDO. Since $O$ is an eigenvector of $\mathcal{D}$, it follows from \cref{eq:SO-theory3} that
\begin{equation}\label{eq:DSO6}
 \<O(0) O(\vec{r})\>_{N,M} = \rho^2 \< O (0) O (\vec{r}/b) \>_{N/b,M/b}.
\end{equation}
Assuming that we can pass to the large volume limit $N,M\to\infty$ in this equation, we obtain 
\begin{equation}\label{eq:DSO7}
	\<O(0) O(\vec{r})\> = \rho^2 \< O (0) O (\vec{r}/b) \>,
\end{equation} 
which means that $O$ is a local scaling operator in the tensor network.

But of course, the large volume and distance limit of correlators of the critical tensor network should be in agreement with the CFT. The two-point correlator of a CFT operator $\mathcal{O}$ scales as $r^{-2\Delta_{\mathcal{O}}}$ and satisfies the discrete scale invariance relation: 
\begin{equation}\label{eq:DSO8}
 \< \mathcal{O}(0) \mathcal{O} (\vec{r}) \>_{\rm CFT} = b^{-2\Delta_{\mathcal{O}}} \< \mathcal{O}(0) \mathcal{O} (\vec{r}/b) \>_{\rm CFT}\,.
\end{equation}
(Similar relations hold for higher-point CFT correlation functions.)
Comparing \eqref{eq:DSO7} and \eqref{eq:DSO8} we obtain precisely the relation \eqref{eq:SO-theory4} for non-rotating RG maps.

A remark is in order concerning the sign of $\rho$. Eq.~\eqref{eq:DSO8} would also be consistent with CFT for $\rho =  - b^{-\Delta_{\mathcal{O}}}$. However as mentioned there is a similar equation for $n$-point correlators (of not necessarily equal operators), which has coefficient $\prod_{i=1}^n\rho_i$ in the r.h.s. Using that equation with $n$ odd and comparing it with CFT, one can show that $\rho$ must be positive for $\mathbb{Z}_2$-even operators. For $\mathbb{Z}_2$-odd operators, odd-$n$ correlators vanish, so this argument only shows that all $\mathbb{Z}_2$-odd operators must have $\rho$ of the same sign. However, a moment's thought shows that, for basically the same reason, $\mathcal{D}$ is defined only up to a sign in the $\mathbb{Z}_2$-odd sector. (This is because when holes surrounded by $\mathcal{D}$ (see \cref{eq:SO-theory3}) are filled with $\mathbb{Z}_2$-odd tensors, one needs an even number of holes in a tensor network with periodic boundary conditions to have a non-vanishing result.) So redefining $\mathcal{D}\to \mathbb{Z}_2 \mathcal{D}$ if necessary, we may make all $\rho$'s positive.

Consider now the case when the RG map is rotating. First, using the LDO definition, for an arbitrary $n$-point function on the tensor network we obtain an equation similar to \cref{eq:DSO7}, but with operators' positions in the r.h.s.~rotated by $\pi/2$ and $\rho^2$ replaced with $\prod_{i=1}^n \rho_i$ (operators are not necessarily equal). Then, we consider the corresponding equation for CFT operators, and match the two. It follows from the transformation laws of conformal operators that the matching requires the relation given in \eqref{eq:SO-theory4} for rotating RG maps.\footnote{As discussed above, $\mathbb{Z}_2$-odd operators have an additional sign ambiguity. We can resolve this by redefining $\mathcal{D} \rightarrow \mathbb{Z}_2 \mathcal{D}$ if needed, ensuring that the leading $\mathbb{Z}_2$-odd eigenvalue is positive (it is real, as it corresponds to the scalar $\sigma$ operator).}

We expect that the correspondence between 8-leg LDO eigenvectors and local CFT operators (primaries or derivative) is going to be one-to-one. So, for any CFT operator $\calO$ we expect that there will be a corresponding $\mathcal{D}$ eigenvector $O$ whose eigenvalue satisfies \cref{eq:SO-theory4}.\footnote{This may appear not so obvious because for lattice models with spins taking a finite number of values, we would need to consider operators of larger and larger range to reproduce infinitely many CFT operators. However, in our case, when tensors are allowed to have infinite bond dimension, we expect that a finite-range representation should be possible for all CFT operators.} Conversely, at least for sufficiently filtering RG maps solving the CDL problem, like Gilt-TNR, we expect that different $\mathcal{D}$ eigenvectors give rise to different CFT operators. This one-to-one correspondence is what we indeed observed in our numerical studies, where the multiplicity of $\mathcal{D}$ eigenvalues (up to numerical errors) always coincided with the expected multiplicity of CFT states. 

\begin{remark}
	We would like to emphasize that our derivation of the LDO spectrum differs substantially from the original proposal in \cite{TNRScaleTrans}. There, an LDO was interpreted as a transfer matrix on a cylinder obtained from the plane by a logarithmic map. The first line of \cref{eq:SO-theory4}\footnote{Ref.~\cite{TNRScaleTrans} did not consider rotating RG maps.} then was argued to follow, since translation on such a cylinder should correspond to dilatation on the plane. One aspect of that argument that is not totally clear to us is which precise observable is being matched between the tensor RG and the CFT. We invite the reader to check the details in \cite{TNRScaleTrans}. We hope our derivation is more direct and allows for a further understanding of the LDO concept. It also makes explicit which observables are being matched (the correlation functions). 
\end{remark}

\subsection{LDO for Gilt-TNR}\label{app:GiltLDO}

As mentioned, in our numerical computations, we used Gilt-TNR \cite{GILT,Gilt-TNR-code,our-code} with minor modifications described in \cite[App.~C]{Ebel:2024nof}. The reader is invited to open \cite[Sec.~III~A]{Ebel:2024nof} for a recap of non-rotating Gilt-TNR. Here we will not repeat that whole discussion, but just recall that the main step of Gilt-TNR is to insert into plaquettes of $A$ tensors specially chosen matrices $Q_1$,$Q_2$,$Q_3$,$Q_4$ which reduce CDL-like correlations around this plaquette. These matrices are then SVD-decomposed:
\beq
\myinclude[scale=0.7]{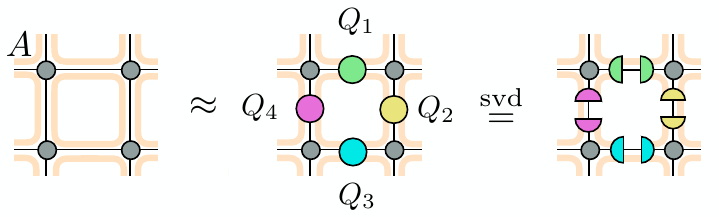}\,,
\label{eq:gilt1}
\eeq
This transformation is performed on every other plaquette in a checkerboard pattern. The subsequent steps are summarized by this equation:
\beq
\myinclude[scale=0.61]{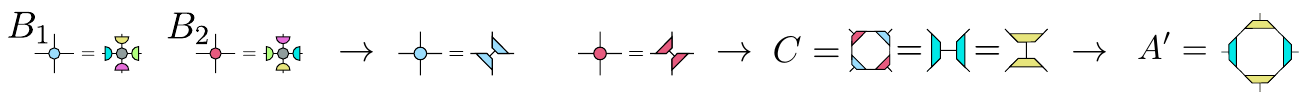}\,.
\label{eq:gilt-summary}
\eeq
Namely SVD halves of matrices $Q_i$ are contracted with tensors $A$ to form tensors $B_1$ and $B_2$. Then, tensors $B_1$ and $B_2$ are SVD-decomposed along the diagonal. The pieces of $B_1$ and $B_2$ are contracted to form a new four-legged tensor $C$ which is also decomposed along its diagonals. Finally, the pieces of the last two SVD decompositions are contracted to form the tensor $A'=\mathcal{R}(A)$.

\cref{fig:DSO1}(a)-(d) show what happens when we apply the same non-rotating Gilt-TNR in the presence of a $2\times 2$ hole in a network made out of $A^*$. Following this series of shown transformations, we obtain the LDO depicted in \cref{fig:DSO1}(e). This is the LDO used in our computations. We should note that the transition from \cref{fig:DSO1}(a) to \cref{fig:DSO1}(b) involves an interesting subtlety, which is discussed in \cref{app:incmpltGilt}.

The notable feature of \cref{fig:DSO1} is that the $2\times 2$ hole shape is preserved. We will be therefore able to diagonalize the LDO and apply the theory developed in \cref{sec:LDOeigs}. It was not guaranteed without trying that the shape would be preserved. In fact, there are two methods for performing the Gilt-TNR step in the presence of a $2\times 2$ hole: the method shown in \cref{fig:DSO1}, and another where the positions of $B_1$ and $B_2$ tensors in \cref{fig:DSO1}(b) are shifted by one in some direction. Carefully following all algorithm steps, one finds that the second method enlarges the hole. 

Finally, for the rotating Gilt-TNR algorithm, the LDO will be given in terms of the non-rotating LDO by \cref{eq:LDOcirc}.

\begin{figure}
	\centering
	\includegraphics[scale=1.2]{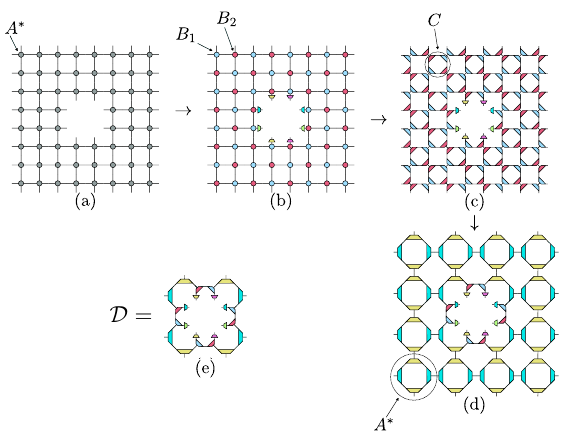}
	\caption{The Gilt-TNR step in the presence of a $2\times 2$ hole. (a) A tensor network with a hole. (b) The network after inserting the Gilt operators $Q_1, Q_2, Q_3, Q_4$. Tensors $B_1, B_2$ are defined in \eqref{eq:gilt-summary}. The semi-circles are the halves of $Q_i$ matrices split by SVD. (c) The network after the first TRG step: tensors $B_1, B_2$ are split into halves using SVD as in \eqref{eq:gilt-summary}. (d) The network after the second TRG step: tensor $C$ (Panel (c)) is split into halves using SVD as in \eqref{eq:gilt-summary}. (e) The LDO for Gilt-TNR.
		   \label{fig:DSO1}
	}
   \end{figure}

\subsection{Numerical results}
\label{sec:DSO-numerics}

Before discussing our numerical results, a comment on the computational cost of the LDO method compared to the TM is in order. As with the TM, we use Arnoldi's method to compute the LDO eigenvalues. The time required to apply the LDO to a tensor scales as $T\sim \chi^8$, so the full computation scales as $O(k \times \chi^8)$, where $k$ is the number of eigenvalues to be computed. Thus, the LDO scaling is comparable to that of the $r=4$ direct and $r=5$ crossed TM (see Eqs.~\eqref{eq:TMscaling}).\footnote{ This partially explains our choice of $r$ for the TM experiments. As noted previously (\cref{foot:r45}), we did not extend the crossed TM computation to $r=5$ because of memory limitations and the excellent results already obtained with the $r=4$ crossed TM.} Let us now proceed to the discussion of the results.

The numerical results for the LDO spectrum can be found in a spreadsheet accompanying this article. The accuracy of the positive scaling dimensions up to $\Delta=4\oneeighth$ is summarized in \cref{tab:LDOerrors}. (For the identity operator, whose exact scaling dimension is $\Delta_{\mathds{1}}=0$, the LDO gives scaling dimension $\approx 2\times10^{-5}$.) These errors can be compared with the errors for the scaling dimensions of quasiprimaries determined from the eigenvalues from the RG map Jacobians, \cref{tab:Jac-errors}, and with the scaling dimensions of quasiprimaries determined from the TM approach, \cref{tab:TMerrorsr4}. Note that the LDO method does not find the central charge. 

\begin{table}[]
	\centering
	\begin{tabular}{llll}
		\toprule
		& $\mathcal{R}$ &  $\mathcal{R}^\circ$ \\
		\midrule
		$\Delta=\oneeighth$ & 0.19\%                           & 0.12\%                         \\
		$\Delta=1$     & 0.07\%                                 & 0.05\%                              \\
                $\Delta=1\,\oneeighth$ & 0.12\%                                 & 0.09\%                      \\
                $\Delta=2$     & 0.18\%                                 & 0.15\%                              \\
                $\Delta=2\,\oneeighth$ & 0.29\%                          & 0.29\%                      \\
                $\Delta=3$     & 0.32\%                                   & 0.26\%                              \\
                $\Delta=3\,\oneeighth$ & 0.46\%                                  & 0.36\%                      \\                
                $\Delta=4$     & 0.39\%                                 & 0.30\%                              \\
                $\Delta=4\,\oneeighth$ & 0.93\%                             & 0.77\%                      \\            
		\bottomrule
	\end{tabular}
	\caption{Relative errors for the scaling dimensions extracted using LDO for the approximate fixed points of the non-rotating map $\mathcal{R}$ and the rotating map $\mathcal{R}^\circ$.
For degenerate scaling dimensions, we show the maximal value of the error. In all cases shown in this table, the number of states and the partially available spin information agree between the LDO computation and the CFT.}
	\label{tab:LDOerrors}
	\end{table}

For the non-rotating map, the LDO computation does not give the spins. For the rotating map the spins can be computed mod $4$ using Eq.~\eqref{eq:SO-theory4}. The resulting spins agree closely with the CFT predictions, with errors $\lesssim 0.01\%$ for all cases in \cref{tab:LDOerrors}. For example, for the six states at $\Delta=3\oneeighth$, the LDO method finds the following six spin values modulo 4:
\beq
\pm 0.99982,
\pm 0.99992,
\pm 1.00007
\label{eq:LDO-spin}
\eeq
which are quite close mod $4$ to the CFT values of $-3,-3,-1,1,3,3$. It is worth noting the contrast with the TM method: while the LDO can only determine spins modulo 4, the TM can, in principle, determine them modulo $r$ (direct TM) and $1+r^2$ (crossed TM) for $r=2,3,4,\ldots$.
 
 As for the scaling dimensions, the performance of the LDO is very similar to that of the $r=4$ TM. This is in spite of the fact that the LDO eigenvalues \cref{sec:LDOeigs} decay for $b=2$ as $0.5^{\Delta}$, i.e.~somewhat slower than the $\approx 0.2^\Delta$ decay for the TM eigenvalues for $r=4$, see Eq.~\eqref{eq:muDelta}. It would be nice if LDO vs TM accuracy comparison could be predicted solely from the eigenvalue decay rate,\footnote{We thank Atsushi Ueda for a discussion.} but as we see this is apparently not the case. It could be that the additional leg compression used in the LDO computation (see \cref{app:compression}), introduced further errors. We also note a subtlety in the interplay between the Gilt algorithm and the LDO, as discussed in \cref{app:incmpltGilt}, which may have also degraded the LDO accuracy. This deserves further study.
 
The similarity of LDO and $r=4$ TM performance extends to the breakdown test, which we performed in the same manner as for the $r=4$ crossed TM at the end of \cref{sec:transfer-numerics}. Specifically, we extended the eigenvalue analysis to levels higher than those reported in \cref{tab:LDOerrors}, focusing on the rotating LDO. We found approximate level quantization, with the number of states in agreement with the CFT, at levels up to $\Delta=6$.  As in the TM case, we use an absolute error exceeding $0.5$ as the criterion for the breakdown of approximate level quantization. This first occurs at the last state on the $\Delta=6 \oneeighth$ level. In total, the 109 states before this breakdown are determined with a maximum absolute (relative) error of 0.21 (3.5\%). This is comparable with our result for $r=4$ crossed TM.

Finally, we computed the scaling dimensions using the LDO for $A_{(0)}$---the less accurate fixed point of $\mathcal{R}^\circ$, with an accuracy of $\delta \sim 3 \times 10^{-3}$. The results are included in the spreadsheet accompanying this article; here, we report only the errors: 
\beq \label{eq:LDOimpr}
\begin{matrix}
	\Delta & \oneeighth & 1 & 1\oneeighth & 2 & 2 \oneeighth & 3 & 3\oneeighth & 4 & 4 \oneeighth \\
	A_{(0)} & 1.30\% & 0.17\% & 0.17\% & 0.21\% & 0.33\% & 0.30\% & 0.39\% & 0.33\% & 0.80\% \\
	A_{(m_*)}  & 0.12\% & 0.05\% & 0.09 \% & 0.15\% & 0.29\% & 0.26\% & 0.36\% & 0.30\% & 0.77\%  \\  
\end{matrix}
\eeq
Here, the first row lists the exact scaling dimensions, while the second and third rows list the maximal errors for the scaling dimensions obtained using $A_{(0)}$ and $A_{(m_*)}$, respectively. The third row corresponds to the last column of \cref{tab:LDOerrors}. The data in \cref{eq:LDOimpr} shows that using a better fixed point approximation yields a significant improvement for the low part of the spectrum, while having little effect on the higher part. A comparison with the TM result (\cref{eq:r4impr}) indicates that the LDO method is somewhat more sensitive to the precision of the fixed point (e.g. three-fold improvement for $\Delta=1$ for LDO vs two-fold improvement in the TM case). However, it still appears that the error is dominated by finite bond dimension effects.               

\section{Conclusions and outlook}
\label{sec:conclusions}
The TM and LDO methods studied in this paper are unique to tensor RG. In some sense, both methods are manifestations of the locality of tensor RG maps. The TM for tensor networks, unlike its well-known counterpart in statistical mechanics, allows the extraction of CFT data using only one tensor, i.e., local network data (the price being that the fixed point tensor is infinite-dimensional). The LDO method has locality at its very core, as the definition of an LDO requires that an RG map can be applied to only a part of the network (outside of the holes). Both methods allow the extraction of CFT scaling dimensions from the fixed point tensor. While the TM method is limited to 2D, the LDO method should work also in 3D and higher dimensions, provided a successful numerical implementation of tensor RG is found in those dimensions. Unlike the Jacobian method used in our previous work \cite{Ebel:2024nof}, these methods successfully determine the scaling dimensions of all CFT operators, and not just quasiprimaries.

After obtaining very precise fixed point tensors in \cite{Ebel:2024nof}, it is of interest to examine how the TM and LDO methods perform when applied to them. This was the purpose of our paper. We also took the opportunity to carefully explain the foundations of the TM and LDO methods, and to advance both by demonstrating how to extract partial information about spins, in addition to the scaling dimensions.

Our numerical findings can be summarized as follows:
  
\begin{itemize}
	\item
	 We found that the LDO method works better than the "narrow" TM method, where TM is built from up to $r=2$ approximate critical tensors. However, wider TM methods, from $r=4$ tensors, matched the LDO accuracy. This was achieved with comparable computational resources and algorithmic scaling with the bond dimension: $\chi^8$ for LDO and the "direct" variant of the $r=4$ TM method, and somewhat faster $\chi^7$  for the "crossed" TM.

      \item We studied the effect of the accuracy of the approximate fixed point on the accuracy of the numerical CFT scaling dimensions. We compared the approximate fixed point of the non-rotating RG map obtained by the shooting method (known to $\delta \sim 5\times 10^{-5}$ accuracy) with the approximate fixed point of the rotating RG map obtained by the Newton method (known to $\delta \sim 10^{-9}$ accuracy). Although the latter fixed point is known to much better accuracy, our numerical results in \cref{tab:TMerrorsr4} and \cref{tab:LDOerrors} show essentially the same accuracy for both the LDO method and the TM method. To further study the effect of the accuracy of the approximate fixed point, we compared the rotating map with a relatively poor approximate fixed point obtained by shooting with an accuracy $\delta \sim 3\times 10^{-3}$ and the highly accurate approximate fixed point obtained by the Newton method. We observed a noticeable reduction in the accuracy of low-lying CFT scaling dimensions when the poor approximate fixed point was used. The LDO method showed a slightly greater sensitivity to the fixed point accuracy in this experiment.

	\end{itemize}

We would like to conclude by listing several open problems:
\begin{itemize}
	\item Given the elegance of the LDO method and its direct relation to the scale transformations, why is it not more widely used? One reason is that, unlike the transfer matrix method, implementing the LDO for a particular RG map requires studying that specific map. In this paper, we detailed the LDO construction for Gilt-TNR. Many other local tensor RG maps also have an LDO preserving a hole of a fixed size \cite{TNRScaleTrans, Evenbly-TNR-website, Evenbly-review, Bal:2017mht}. However, it is not clear if the existence of LDO is guaranteed in general. Understanding the conditions that guarantee the existence of an LDO would be a very interesting accomplishment. 
	\item The eigenstates obtained in the LDO method and the Jacobian method (for eigenvalues of quasiprimaries) correspond to local operators with well-defined scaling dimensions. Is there perhaps a relationship between the eigenstates of the LDO and the eigenstates of the Jacobian? 
	\item In this paper, we focused on CFT scaling dimensions and spins. Another important part of the data characterizing the CFT are the operator product expansion (OPE) coefficients. In the past, several works have shown that the OPE coefficients can also be extracted from the tensor RG fixed point tensor \cite{Evenbly:2017dyd,PhysRevE.109.034111,TNRScaleTrans,Li_2022,Ueda_2023}. Will the accuracy in determining the OPE coefficients improve by using the high-quality fixed point? 
	
	\item A much more open-ended problem is as follows. In \cite{Ebel:2024nof} and especially in this work we provided a lot of evidence that, in the case we studied, tensor RG fixed point properties are compatible with the underlying CFT. However, only a minuscule part of CFT spatial symmetries was manifest in the tensor network description. For example, we demonstrated that LDO provides an explicit construction for scaling operators in the tensor network, but only discrete scale invariance was manifest, while of course CFT has a continuous scale symmetry, not to speak of conformal invariance. Is there a way to argue that, under certain conditions, fixed points of tensor RG are guaranteed to correspond to fully rotationally and translationally symmetric field theories, even though those spatial symmetries are not manifest?

\end{itemize}

\begin{acknowledgments}

This research was supported in part by the Simons Foundation grant 733758 (Simons Bootstrap Collaboration). This research was supported in part by grant no. NSF PHY-2309135 to the Kavli Institute for Theoretical Physics (KITP). We thank Cl\'ement Delcamp and Atsushi Ueda for discussions.

\end{acknowledgments}

\appendix

\section{Transfer matrix details}
\label{app:TM}

Here we give a derivation of Eqs.~\eqref{eq:TMeigs1},\eqref{eq:Teigs},\eqref{eq:TMeigs2} for the TM and translation generator eigenvalues. We will consider the case $b=2$, the general case being essentially identical.

Consider first the direct TM for general $r$. Take the $N\times M$ tensor network where $N=r 2^n$, $M=s 2^n$, $s\in \mathbb{N}$, built out of the critical fixed point tensor $A^*$ which is the fixed point of the non-normalized RG map. For $n\gg 1$, its partition function must be equal to the CFT partition function on the $N\times M$ torus, allowing for a volume-proportional term in the free energy which the CFT does not predict. I.e.~we expect
\beq
Z(A^*, N\times M)\approx e^{-f N M} Z_{\rm CFT}(N\times M)\,\qquad(N,M\gg 1),
\label{eq:TMapp1}
\eeq
where
\beq
Z_{\rm CFT}(N\times M)= \sum_{k=0}^\infty e^{- 2\pi \frac{s}{r} (\Delta_k-c/12)}.
\eeq
Now let's compute $Z(A^*, N\times M)$ using RG. Assume first that the map is non-rotating. We have (for an RG map with $b=2$)
\begin{align}
	Z(A^*, N\times M) &= Z(A^*, \frac N2\times \frac M2)=\ldots=Z(A^*, r\times s)\,,
	\label{eq:TMapp2}
\end{align}
where in $\ldots$ we keep applying RG until we get to a network of size $r\times s$.
The last term here can be expressed in terms of the direct TM $\calM_r$ eigenvalues $\mu_k$ :
\beq
\sum_{k=0}^\infty (\mu_k )^{s}\,\label{eq:TMapp3}
\eeq
Using \eqref{eq:TMapp1}, \eqref{eq:TMapp2}, \eqref{eq:TMapp3}, we conclude that $f=0$, while $\mu_k$ are related to the CFT scaling dimensions via \eqref{eq:TMeigs1}.

Next let us discuss the eigenvalues of the translation generator $\calT$. We consider the $N\times M$ tensor network built out of $A^*$, with the same $N$ and $M$ as above, but we change the boundary conditions. We consider periodic boundary conditions in the horizontal direction, and twisted periodic boundary conditions in the vertical direction, where lower outgoing leg $i$ is contracted with upper outgoing leg $i+M$ modulo $N$:
\beq
\myinclude[scale=1]{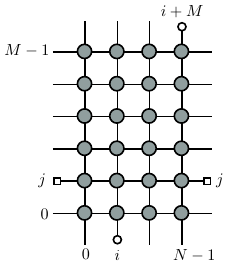}\,.
\eeq
In the large tensor network size limit, this corresponds to the CFT partition function on the torus of modular parameter $\tau=(1+i)s/r$:
\beq
\myinclude[scale=1]{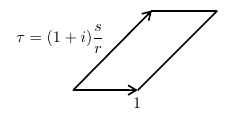}\,,
\eeq
which can be computed using the general formula \cite{DiFrancesco:1997nk}
\beq
Z_{\rm CFT}(\tau)= \sum_{k=0}^\infty e^{- 2\pi\, {\rm Im}\tau\, (\Delta_k-c/12)+2\pi i\, {\rm Re}\tau\, \ell_k}\,.
\eeq
On the other hand, applying the RG map $n$ times, we can reduce the tensor network to size $r\times s$. Here we use the assumption that the RG map is $(h,0)$-twist friendly with $h|2$. We assume that the RG map acts locally and can be performed in the presence of twisted boundary conditions just as for pure periodic boundary conditions. This holds for all reasonable RG maps and in particular for Gilt-TNR. The partition function of the $r\times s$ tensor network with twisted boundary conditions is given by \eqref{eq:TMapp3} where $\mu_k$ are the eigenvalues of $\calT\calM_r$. Comparing with the CFT partition function, and using the already determined eigenvalues of $\calM_r$, we conclude that the eigenvalues of $\calT$ are given by $e^{i \frac{2\pi}r \ell_k}$.

Finally let us compute the eigenvalues of the crossed TM. For this we consider the $N\times M$ tensor network built out of $A^*$, with the same $N$ and $M$ as above, but with yet another choice of boundary conditions. This time consider periodic boundary conditions in the vertical direction, and twisted periodic boundary conditions in the horizontal direction, where the left outgoing leg $j$ is contracted with the right outgoing leg $j+2^n$ modulo $M$:
\beq
\myinclude[scale=1]{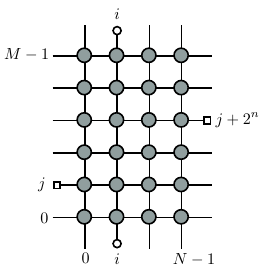}\,.
\eeq
We will consider $s\ge 2$ in $M=s 2^n$, because for $s=1$ this partition function is just the usual partition function with the periodic boundary conditions. In the large tensor network size limit, this twisted partition function corresponds to the CFT partition function on the torus of the shape
\beq
\myinclude[scale=1]{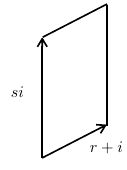}\,,
\eeq
whose modular parameter is given by $\tau=si/(r+i)=\frac{s}{r^2+1}+i \frac{s r}{r^2+1}$. We apply the RG map $n$ times, reducing the tensor network to size $r\times s$. Here we use the assumption that the RG map is $(0,v)$-twist friendly for $v|2$. With the current boundary conditions, the resulting partition function is given by \eqref{eq:TMapp3} where $\mu_k$ are the eigenvalues of $\widetilde\calM_r$. Comparing with the CFT partition function, we deduce \eqref{eq:TMeigs2}.

If the RG map is rotating, the above reasoning still applies. We just have to choose $n$ in $N=r 2^n$ a multiple of 4. Then, after $n$ RG steps, the lattice orientation comes back to itself. This implies that the equations for TM eigenvalues remain valid.

Given an exact fixed point tensor $A^*$, Eqs.~\eqref{eq:TMeigs1}, \eqref{eq:TMeigs2} should agree with the CFT spectrum and the central charge, no matter which $r$ is used. In our numerical tests, involving an approximate fixed point tensor of a finite bond dimension $\chi=30$, we used $r=1,2,3$. Looking at the states with $\Delta \le 4$, we observed significant improvement for higher-lying states when moving from $r=1$ to $r=2$, and further improvement for $r=3$. The spectrum results reported in \cref{sec:transfer-numerics} use $r=2$.

The improvement for higher $r$ can be rationalized as follows. Controlling higher scaling dimensions is tantamount to controlling the partition function for $M/N\gg 1$. It is natural to expect that this may require using a larger bond dimension in the vertical ($M$) direction than in the $N$ direction. And using the transfer matrix based on $r$ copies of $A^*$ effectively increases the vertical bond dimension from $\chi$ to $\chi^r$.

\section{Lattice dilatation operator details}
\label{app:ScalOp}

In this appendix, we provide details on the LDO used to compute the scaling dimensions for \cref{tab:LDOerrors}. In Section~\ref{app:incmpltGilt}, we explain a subtlety occurring when we insert Gilt matrices into incomplete plaquettes in the transition between panels (a) and (b) in \cref{fig:DSO1}. In Section~\ref{app:compression}, we discuss the memory requirements of the LDO computation and describe the approach (given in "Evaluation of scaling dimensions" code in \cite{Evenbly-TNR-website}) that significantly decreases the memory cost. 

\subsection{Gilt error in incomplete plaquettes}\label{app:incmpltGilt}

Let us examine the transition between (a) and (b) in \cref{fig:DSO1} more closely. At first glance, there appears to be an (approximate) equality between (a) and (b). We inserted Gilt matrices $Q_1, Q_2, Q_3, Q_4$ into plaquettes of the network, \cref{eq:gilt1}, which by design of the algorithm is meant to introduce only a small truncation error. This step is followed by absorption of pieces of Gilt matrices into $B_1$ and $B_2$ as shown in \eqref{eq:gilt-summary}. Thus, only the non-absorbed halves of Gilt matrices on the legs around the hole appear in \cref{fig:DSO1}(b).

However, upon closer inspection, we see that the passage from (a) and (b) is not obviously legal, as we inserted Gilt matrices into incomplete plaquettes, i.e.~plaquettes with one tensor missing, in the corners of the hole. Thus, we performed the following replacement for the upper left corner:
\begin{equation}\label{eq:Ptrick1}
   \myinclude{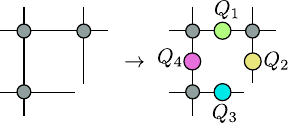}\,,
\end{equation}
and similar replacements for the other three corners. This replacement is more general than what \cref{eq:gilt1} says, and one may wonder if we are allowed to do this.
 
We can do some tests of this replacement. The first test would be to compare the l.h.s.~and r.h.s.~of \cref{eq:Ptrick1}. The results of this test are not very encouraging. E.g., using the approximate critical fixed point tensor $A^{(n_*)}$ (see \cref{fig:crit_convergence} and the subsequent discussion), the relative difference\footnote{I.e. $\|X-Y\|/\|X\|$ where $X$ and $Y$ are the two tensors.} between l.h.s. and r.h.s. of \cref{eq:Ptrick1} comes out $\sim 0.25$. This may suggest that the replacement \eqref{eq:Ptrick1} introduces a huge error in the network. This is puzzling, as we have seen that the LDO derived through this replacement reproduces CFT scaling dimensions with excellent accuracy (\cref{tab:LDOerrors}). What's going on?

The second test, which is more to the point, is as follows. The point is that we are not interested in comparing l.h.s. and r.h.s. of \cref{eq:Ptrick1} for all indices in the lower left corner, but only for those pairs of these indices which are relevant for the computation of the leading LDO eigenvectors. In other words, the replacement we are interested in is 
\begin{equation}\label{eq:test2}
	\myinclude{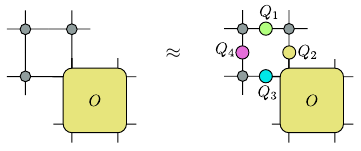}\,,
\end{equation} 
where $O$ is one of the leading LDO eigenvectors.  It turns out that this test works out perfectly. For all non-rotating LDO eigenvectors whose eigenvalue errors are reported in \cref{tab:LDOerrors}, the relative error between the two sides of \eqref{eq:test2} is $\lesssim 10^{-4}$, comparable to the error that the Gilt algorithm introduces in the full plaquette \cref{eq:gilt1}.\footnote{We believe that the same should also hold for the rotating LDO eigenvectors, although numerical checks were not performed in that case.}

This can be explained based on the form of our LDO. 
We know that if $A^*$, given by (see \eqref{eq:gilt-summary})
\beq
\myinclude[scale=1.5]{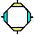}\,,
\eeq
is glued into \eqref{eq:Ptrick1}, the error is small. Now, we have checked that the error remains small if the following part of $A^*$ is glued in: 
\beq
\label{eq:Apart}
\myinclude[scale=1.2]{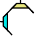}\,.
\eeq
In other words,
\begin{equation}\label{eq:test3}
	\myinclude[scale=0.8]{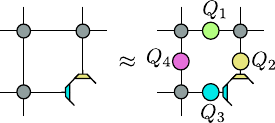}\,.
\end{equation} 
Numerically, we see that the relative error in \cref{eq:test3} is of order $10^{-3}$.

This implies that \eqref{eq:test2} is guaranteed to hold with a small error for any $O$ which is an eigenstate of LDO. For the proof, replace both occurrences of $O$ in \eqref{eq:test2} by $\mathcal{D}O$ and note that \eqref{eq:Apart} occurs in the upper left corner of our LDO, \cref{fig:DSO1}(e)

\subsection{Leg compression}\label{app:compression}

Let us discuss one crucial technical detail important for performing numerical diagonalization of $\mathcal{D}$. We use 64-bit floating point numbers to represent tensor elements. An eigenvector of $\mathcal{D}$ is an eight-legged tensor, and for the bond dimension $\chi=30$ and $64$ bits per tensor element requires around $5$ terabytes of memory. For this computation we used a computer with about 100 gigabytes of RAM, on which we could not store even a single eight-legged tensor, let alone use Krylov's method to find the spectrum of $\mathcal{D}$. To reduce the memory cost and make it manageable, we used instead a method from the "Evaluation of scaling dimensions" code in \cite{Evenbly-TNR-website}, which we will now describe.

We start by performing a projective truncation \cite{Evenbly-review} of the network with a hole as follows (relative length changes of some network bonds are for drawing purposes only):
\begin{equation}\label{eq:DS10}
 \myinclude[scale=0.8]{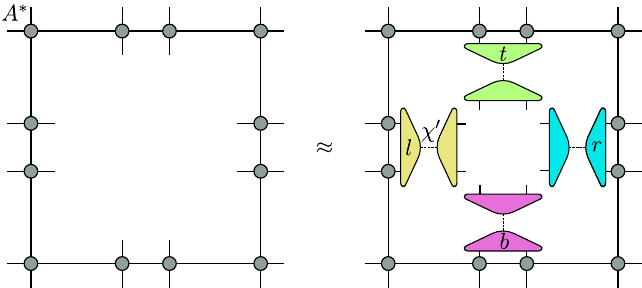}\,. 
\end{equation}
Here, $l,r,t,b$ are isometries, and the bond dimension $\chi'$ is the parameter controlling the accuracy. The method for finding $l,r,t,b$ will be discussed at the end of this section. Next, we apply Gilt-TNR to the r.h.s. of \cref{eq:DS10}:
\begin{equation}\label{eq:DS105}
 \myinclude[scale=0.8]{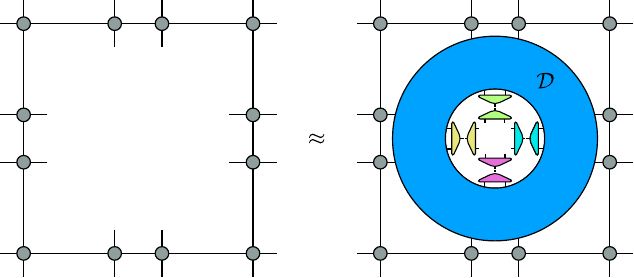}\,. 
\end{equation} 
Here, the blue circle represents the Gilt-TNR $\mathcal{D}$ tensor given by \cref{fig:DSO1}(e). Finally, we again apply the projective truncation from \cref{eq:DS10}, this time to the r.h.s. of \cref{eq:DS105}. This results in:
\begin{equation}\label{eq:DS11}
 \myinclude[scale=0.8]{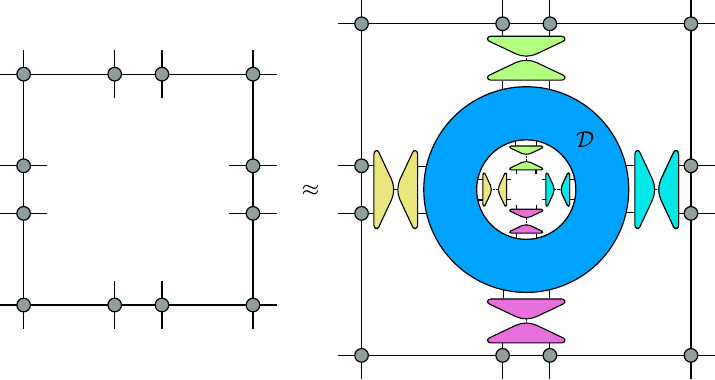}\,.
\end{equation} 
\cref{eq:DS11} shows that $\mathcal{D}$ surrounded by projectors $l l^t, r r^t, t t^t,$ and $b b^t$ approximately solves \cref{eq:SO-theory3} with respect to $\mathcal{D}$. Then, as we showed in \cref{sec:DSO-theory}, its spectrum should satisfy \cref{eq:SO-theory4}.\footnote{Note that for $\chi'=\chi^2$, Eqs.~\eqref{eq:DS10}-\eqref{eq:DS11} become exact (up to the errors of the Gilt-TNR algorithm itself) as all projectors become identity matrices.} 

On the other hand, the nonzero part of the spectrum of $\mathcal{D}$ surrounded by projectors as in \cref{eq:DS11} coincides with the spectrum of a truncated operator $\mathcal{D}_{\text{trnc}}$ defined as 
\begin{equation}\label{eq:DS12}
 \myinclude{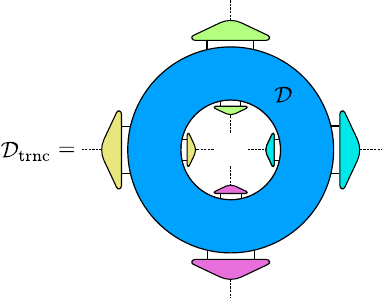}\,.
\end{equation} 
The spectrum of $\mathcal{D}_{\text{trnc}}$ is what we compute in practice.

In our computation, we used isometries $l, r, t, b$ that are obtained by performing truncated SVD on the following environments:
\begin{equation}\label{eq:DS13}
 \myinclude{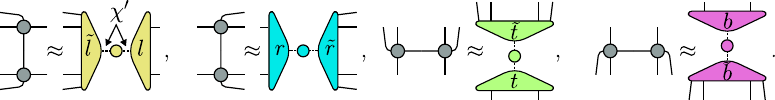}
\end{equation}
Here, the circles are diagonal matrices with the singular values of the corresponding environments on the diagonal. Note that since $\tilde{l}, \tilde{r}, \tilde{t}, \tilde{b}$ are not needed in the computation, instead of performing SVD of these environments, we can use a cheaper eigenvalue decomposition of those multiplied by hermitian conjugates. The scaling dimensions for \cref{tab:LDOerrors} were obtained using $\chi'=40$. Such a computation requires around 90 gigabytes of RAM.

\section{LDO and Jacobian eigenvalues in Refs.~\cite{Lyu:2021qlw,PhysRevE.109.034111}}\label{app:frozenJacs} 

This appendix stands apart from the main line of development of our work. Its main purpose is to shed light on the surprising agreement between Jacobian eigenvalues for the ``frozen'' RG maps defined in \cite{Lyu:2021qlw,PhysRevE.109.034111} and the CFT scaling dimensions of derivative operators. This agreement presents an exception to the general rule of thumb that the derivative operator eigenvalues are not universal in RG, see \cite[Sec.~II~B]{Ebel:2024nof}. As already mentioned in \cite{Ebel:2024nof}, the explanation lies in the fact that the Jacobian of the ``frozen'' RG map in \cite{Lyu:2021qlw,PhysRevE.109.034111} agrees with the LDO. If our explanation is correct,  the agreement should not hold for the eigenvalues of the non-frozen RG map (\cite{Lyu:2021qlw,PhysRevE.109.034111} do not report those eigenvalues).

\subsection{Hole-shrinking LDOs}\label{app:frozenJacs1}
To make contact with \cite{Lyu:2021qlw,PhysRevE.109.034111}, it will be convenient to introduce a slight modification of the LDO concept discussed in \cref{sec:DSO}, namely hole-shrinking LDOs. 

Consider a locally defined non-normalized, non-rotating RG map $\mathcal{R}$ with scaling factor $b$ and a critical fixed point tensor $A^*$: $\mathcal{R}(A^*)=A^*$. Consider a network built out of $A^*$ with periodic boundary conditions and a $2 \times 2$ hole in it. Assume that applying $\mathcal{R}$ to the network, we can transform it into another one, with a $1 \times 1$ hole connected to the original internal $2\times2$ hole legs via a linear operator $\mathcal{D}$:
\begin{equation}\label{eq:DSO10}
\myinclude{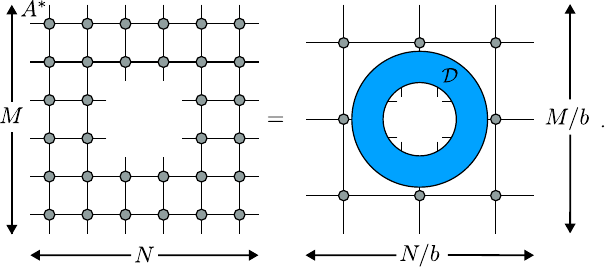}
\end{equation}
When \cref{eq:DSO10} holds, we call $\mathcal{D}$ a hole-shrinking LDO. As in \cref{sec:DSO-theory}, we will assume that if we create multiple holes in the network, then to obtain the same state after an RG step, we should surround each hole with $\mathcal{D}$ operator as in \cref{eq:DSO10}.

To extract scaling dimensions from $\mathcal{D}$, we note that \cref{eq:DSO10} also implies that there is $\mathcal{D}_{1 \times 1}$ satisfying \cref{eq:SO-theory3} with the $2 \times 2$ holes on the r.h.s. and l.h.s. replaced by $1 \times 1$ holes. One can see this by contracting both sides of \cref{eq:DSO10} with three copies of $A^*$ tensor, filling the hole so that only one vacant place is left. The spectrum of $\mathcal{D}_{1 \times 1}$ satisfies \cref{eq:SO-theory4}.\footnote{The proof of this fact is analogous to that for the case of $2\times 2$ holes in \cref{sec:DSO-theory}.} Below, we will describe a slight modification of this idea that will help us to understand the results of Refs.~\cite{Lyu:2021qlw,PhysRevE.109.034111}. The modification consists in averaging over four different ways to fill in the hole.

Let $o$ be a four-legged tensor (we use lowercase $o$ to distinguish it from the eight-legged $O$ in the previous section). Consider the following sum of correlation functions:
\begin{equation}\label{eq:DSO11}
 \myinclude{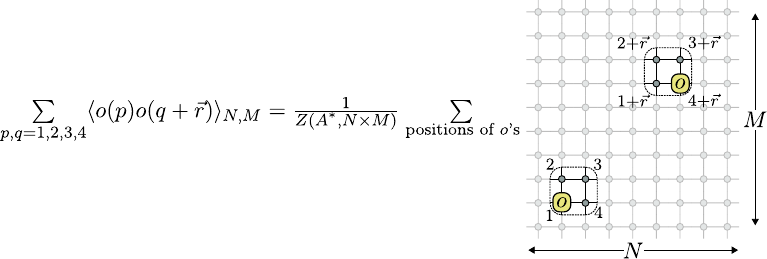}\,.
\end{equation}
Here, $p,q=1,2,3,4$ denote the positions of tensors within $2 \times 2$ blocks as indicated on the r.h.s. of \cref{eq:DSO11}. The sum on the r.h.s. is identical to that on the l.h.s. It runs over the positions of $o$ tensors such that each $2 \times 2$ block, surrounded by a dashed line, contains one $o$ tensor.

Applying an RG step to \cref{eq:DSO11}, we obtain: 
\begin{equation}\label{eq:DSO12}
 \myinclude{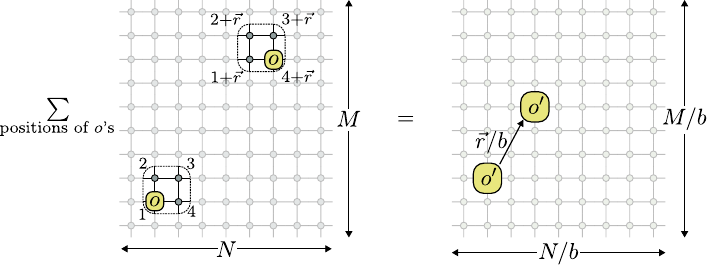}\, ,
\end{equation}
where
\begin{equation}\label{eq:DSO13}
 \myinclude{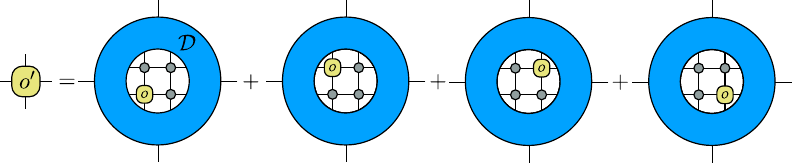}\,.
\end{equation}
We can express \cref{eq:DSO12} in terms of correlation functions:
\begin{equation}\label{eq:DSO14}
 \sum_{i,j=1,2,3,4} \<o(i)o(j+\vec{r}) \>_{N,M} = \<o'(0)o'(\vec{r}/b)\>_{N/b,M/b}.
\end{equation} 

We will now relate the spectrum of the CFT corresponding to $A^*$ with the spectrum of the $\tilde{\mathcal{D}}$ operator defined as follows:
\begin{equation}
 \myinclude{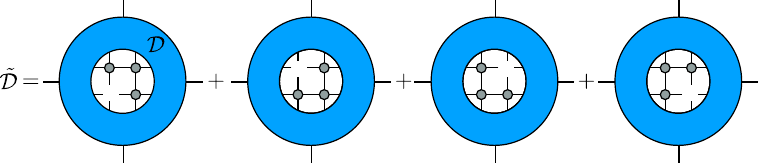}\,. \label{eq:DSO14b}
\end{equation} 
Assume $o$ is an eigenvector of $\tilde{\mathcal{D}}$ with eigenvalue $\rho$. Then, \cref{eq:DSO14} becomes
\begin{equation}\label{eq:DSO15}
 \sum_{i,j=1,2,3,4} \<o(i)o(j+\vec{r}) \>_{N,M} = \rho^2 \<o(0)o(\vec{r}/b)\>_{N/b,M/b}.
\end{equation}
The l.h.s. of \cref{eq:DSO15} has the following asymptotics:
\begin{equation}\label{eq:DSO16}
 \lim_{N,M \rightarrow \infty }\sum_{i,j=1,2,3,4} \<o(i)o(j+\vec{r}) \>_{N,M} \sim_{r \rightarrow \infty} 16 \lim_{N,M \rightarrow \infty } \< o(o) o(\vec{r}) \>_{N,M}.
\end{equation} 
Substituting \cref{eq:DSO16} into \cref{eq:DSO15}, we obtain:
\begin{equation}\label{eq:DSO17}
 \lim_{N,M \rightarrow \infty } \< o(o) o(\vec{r}) \>_{N,M} \sim_{r \rightarrow \infty} \left(\frac{\rho}{4}\right)^2 \lim_{N,M \rightarrow \infty} \<o(0)o(\vec{r}/b)\>_{N,M}.
\end{equation}
On the CFT side, there is an operator $\mathcal{O}$ whose two-point function agrees with that of $o$ in the large volume and $r \rightarrow \infty$ limit. Then, \cref{eq:DSO17} implies that
\begin{equation}\label{eq:DSO18}
 \< \mathcal{O}(0) \mathcal{O}(\vec{r}) \>_{\rm CFT}= \left(\frac{\rho}{4}\right)^2\< \mathcal{O}(0) \mathcal{O}(\vec{r}/b) \>_{\rm CFT}.
\end{equation} 
Therefore, $\rho$ and the scaling dimension $\Delta{\mathcal{O}}$ of $\mathcal{O}$ must satisfy\footnote{See the discussion about the sign ambiguity after \cref{eq:DSO8}.}
\begin{equation}\label{eq:DSO19}
 \rho = 4 b^{-\Delta_{\mathcal{O}}}.
\end{equation}
Using \cref{eq:DSO19}, we can retrieve the CFT spectrum from that of $\tilde{\mathcal{D}}$. Let us now connect this result to those of Refs.~\cite{Lyu:2021qlw,PhysRevE.109.034111}.

\subsection{Connection to Refs.~\cite{Lyu:2021qlw,PhysRevE.109.034111}}\label{app:frozenJacs2} 
Refs.~\cite{Lyu:2021qlw,PhysRevE.109.034111} both use Gilt followed by a HOTRG step \cite{HOTRG} (Gilt-HOTRG) as the RG map.\footnote{See \cite{Lyu:2021qlw,PhysRevE.109.034111}, and a short review in \cite[App.~C.4]{Ebel:2024nof}.} There are some differences in their approaches. The major one is that Ref.~\cite{PhysRevE.109.034111} exploits the minimal canonical form \cite{Acuaviva:2022lnc} of a tensor to fix the gauge. Notably, this helped them obtain a better critical fixed point tensor than that in \cite{Lyu:2021qlw}.\footnote{We note that, to our knowledge, Ref.~\cite{PhysRevE.109.034111} is the first work where the concept of minimal canonical form was applied in practice.}  What is similar in these works is that the reported CFT spectra were obtained by differentiating an RG map with all Gilt matrices, isometries, and gauge transformations fixed to constant tensors in the neighborhood of $A^*$.\footnote{In \cite{Lyu:2021qlw}, this can be inferred from Eq.~(40). In \cite{PhysRevE.109.034111}, this is stated explicitly after Eq.~(18).} As in \cite{Ebel:2024nof}, we call such an RG map ``frozen.''

Here, we will show that the Jacobian of the frozen Gilt-HOTRG is approximately equal to the LDO arising from Gilt-HOTRG in the presence of a $2 \times 2$ hole. We will focus on the map from Ref.~\cite{Lyu:2021qlw}. However, one may make a similar observation for \cite{PhysRevE.109.034111}. This coincidence of the frozen map Jacobian and the LDO explains why numerical spectra in \cite{Lyu:2021qlw,PhysRevE.109.034111} coincide with the CFT prediction even for total derivative operators.    

Let $\mathcal{R}$ be the Gilt-HOTRG map (see \cite[App~C.4]{Ebel:2024nof} and \cite{Lyu:2021qlw}). Applying $\mathcal{R}$ to the network with a $2 \times 2$ hole shows that $\mathcal{R}$ satisfies \cref{eq:DSO10} with $\mathcal{D}$ given by [figure adapted from Ref.~\cite{Lyu:2021qlw}, Eq.~31]:\footnote{A numerical study akin to that in \cref{app:incmpltGilt} is needed to confirm the validity of such an LDO.}
\begin{equation}\label{eq:DSO20}
 \myinclude{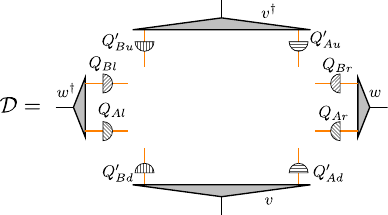}\,.
\end{equation} 
Thus, $\mathcal{D}$ shrinks holes and, according to the discussion in \cref{app:frozenJacs1}, we can extract the scaling dimensions by studying the spectrum of $\tilde{\mathcal{D}}$ given by [figure adapted from Ref.~\cite{Lyu:2021qlw}, Eq.~31]:
\begin{equation}\label{eq:DSO21}
 \myinclude[scale=0.8]{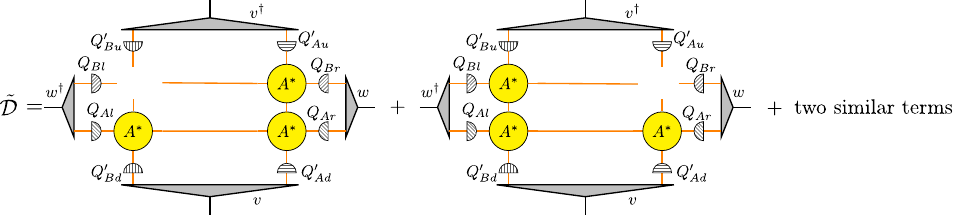}\,. 
\end{equation}  

Ref.~\cite{Lyu:2021qlw} did not extract scaling dimensions using the LDO method discussed here. Instead, they used the linearized RG transformation method. Furthermore, as mentioned earlier, they considered a frozen version of the RG map $\mathcal{R}$ when evaluating the Jacobian. For the frozen map, the Jacobian $\nabla \mathcal{R}$ is given by the expression [figure adapted from Ref.~\cite{Lyu:2021qlw}, Eq.~40]:
\begin{equation}\label{eq:DSO22}
 \myinclude[scale=0.8]{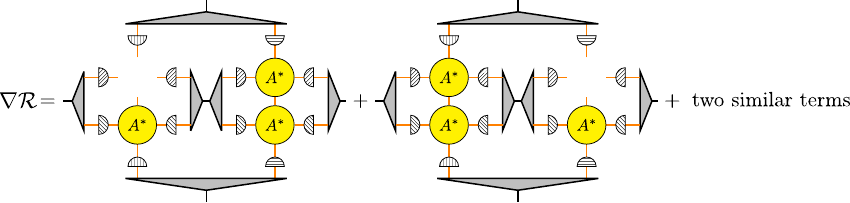}\,.
\end{equation} 
Now, looking at \eqref{eq:DSO22} and \eqref{eq:DSO21}, we see that the two expressions are very similar. The only difference is the presence of additional isometries and halves of $Q$ matrices in the middle of $\nabla \mathcal{R}$. We expect that these network components mainly filter local correlations and do not significantly affect the result of contraction. This expectation, however, should be supported by a numerical study similar to what we presented in \cref{app:incmpltGilt}. Here, we simply assume that our expectation holds. We then conclude that:
\begin{equation}\label{eq:DSO23}
 \nabla \mathcal{R} \approx \tilde{\mathcal{D}}. 
\end{equation}

\cref{eq:DSO23} implies that the spectrum of $\nabla \mathcal{R}$ satisfies \cref{eq:DSO19}. Substituting the scaling factor $b=2$ to \cref{eq:DSO19}, we obtain the relationship between eigenvalue $\mu$ of $\nabla \mathcal{R}$ and the scaling dimension of the corresponding operator $\mathcal{O}$ in the underlying CFT: 
\begin{equation}\label{eq:DSO24}
 \mu = 2^{2-\Delta_{\mathcal{O}}}.
\end{equation}  
Note in this respect that tensor $A^*$ is by construction an eigenvector of $\tilde{\mathcal{D}}$ with eigenvalue $4$; it corresponds to the unit CFT operator.

\cref{eq:DSO24} is exactly the formula used in \cite{Lyu:2021qlw}. We emphasize that \cref{eq:DSO24} holds for any operators $\mathcal{O}$, including total derivatives. This explains the surprising agreement of scaling dimensions obtained in Ref.~\cite{Lyu:2021qlw} with the exact values for total derivative operators.


\providecommand{\href}[2]{#2}\begingroup\raggedright\endgroup

\end{document}